\documentclass[10pt,twocolumn,showpacs,amsmath,amssymb,floatfix,superscriptaddress]{revtex4-2}

\usepackage[export]{adjustbox}
\usepackage{color}
\usepackage[cmyk]{xcolor}
\usepackage{overpic}
\usepackage{graphicx}
\usepackage{dcolumn}
\usepackage{bm}
\usepackage{color}
\usepackage{txfonts}
\usepackage{microtype}
\usepackage{hyperref}
\usepackage[english]{babel}
\usepackage{slashed}
\usepackage{gensymb}
\usepackage{epsfig}
\usepackage[normalem]{ulem}
\usepackage{amsmath}
\usepackage{array}
\usepackage{booktabs}
\usepackage{multirow}
\usepackage{siunitx}
\allowdisplaybreaks[4]

\usepackage[mathscr]{eucal}

\begin{document}
\renewcommand{\figurename}{FIG.}

\title{Generation of Ultra-Collimated Polarized Attosecond $\gamma-$Rays via Beam Instabilities}

\author{Li-Jie Cui}  	\affiliation{Ministry of Education Key Laboratory for Nonequilibrium Synthesis and Modulation of Condensed Matter, Shaanxi Province Key Laboratory of Quantum Information and Quantum Optoelectronic Devices, School of Physics, Xi'an Jiaotong University, Xi'an 710049, China}

\author{Ke-Jia Wei}  	\affiliation{Ministry of Education Key Laboratory for Nonequilibrium Synthesis and Modulation of Condensed Matter, Shaanxi Province Key Laboratory of Quantum Information and Quantum Optoelectronic Devices, School of Physics, Xi'an Jiaotong University, Xi'an 710049, China}

\author{Chong Lv}  	\affiliation{Department of Nuclear Physics, China Institute of Atomic Energy, P.O. Box 275(7), Beijing 102413, China}

\author{Feng Wan}\email{wanfeng@xjtu.edu.cn}
\affiliation{Ministry of Education Key Laboratory for Nonequilibrium Synthesis and Modulation of Condensed Matter, Shaanxi Province Key Laboratory of Quantum Information and Quantum Optoelectronic Devices, School of Physics, Xi'an Jiaotong University, Xi'an 710049, China}

\author{Yousef I. Salamin}
\affiliation{Department of Physics, American University of Sharjah, Sharjah, POB 26666 Sharjah,  United Arab Emirates}

\author{Lei-Feng Cao}\email{caoleifeng@sztu.edu.cn}
\affiliation{Shenzhen Key Laboratory of Ultraintense Laser and Advanced Material Technology, Center for Intense Laser Application 
Technology, and College of Engineering Physics, Shenzhen Technology University, Shenzhen 518118, People’s Republic of China}

\author{Jian-Xing Li}\email{jianxing@xjtu.edu.cn}
\affiliation{Ministry of Education Key Laboratory for Nonequilibrium Synthesis and Modulation of Condensed Matter, Shaanxi Province Key Laboratory of Quantum Information and Quantum Optoelectronic Devices, School of Physics, Xi'an Jiaotong University, Xi'an 710049, China}
\affiliation{Department of Nuclear Physics, China Institute of Atomic Energy, P.O. Box 275(7), Beijing 102413, China}

\date{\today}
	
\begin{abstract}
Polarized attosecond  $\gamma-$rays may offer excitation and hyperfine tracking of reactions relevant to nuclear physics, astrophysics, high-energy physics, etc. However, unfortunately, generation of a feasible and easy-to-deploy source is still a great challenge.
Here, we put forward a novel method for producing ultra-collimated high-brilliance polarized attosecond $\gamma-$rays via the interaction of an unpolarized electron beam with a solid-density plasma. 
As a relativistic electron beam enters a solid-density plasma, it can be modulated into high-density clusters via the self-modulation instability of itself and further into attosecond slices due to its own hosing instability. 
This is accompanied by the generation of similar pulse-width $\gamma-$slices via nonlinear Compton scattering.  
The severe hosing instability breaks the symmetry of the excited electromagnetic fields, resulting in net linear polarization of $\gamma-$slices, which challenges the conventional perception that the interaction of an axially symmetric unpolarized electron beam with a uniform plasma cannot generate polarized radiation. 
In addition, we also obtain high-quality electron microbunches which may serve as an alternative source for prebunched free-electron lasers.
\end{abstract}
	
\maketitle
Exploration of fast-evolving microscopic processes requires pulses with ultra-high temporal resolution to excite and take snapshots \cite{villeneuve2018Atto}.
For example, it takes several femtoseconds to track the destruction and formation of chemical bonds as well as  the motion of atoms and molecules \cite{zewail2000femtochemistry}. 
Finer processes, such as electronic motion in the atomic shells, require time-resolution techniques on the attosecond scale \cite{corkum2007attosecond, Kling2008Atto}.
 For most reaction processes, the excitation energy is in the order of tens of eV or even lower \cite{drescher2002time}, which can be fulfilled with currently available pulses with photon energy in the regime of extreme ultraviolet (EUV) and X-rays \cite{baltuvska2003attosecond_control,Markus2001X-ray}. 
However, for some more complex processes, such as photonuclear reactions and astrophysical processes, the required photon energy for the pulse is in the order of MeV and even higher \cite{utsunomiya2003cross,xu2007new,weller1992angular}. 
Cross-sections of these reactions are typically very small, ranging from a few to tens of millibarns \cite{Photonuclear-data,kossov2002approximation}. 
So, polarization of the source is crucial to enhance the excitation of these reactions \cite{J.Speth_1981Giant,Akbar2017Measurement}. 
Besides, the excitation processes of these reactions typically occur on extremely short timescales \cite{Krausz2009Atto}, such as  femtosecond-scale in nuclear single-particle transition \cite{zilges2022photonuclear} and resonance fluorescence, attosecond-scale in resonance internal conversion, and zeptosecond-scale evolution of compound nuclei \cite{li2015attosecond,povh1995particles}. 
Therefore, attosecond $\gamma-$ray pulses with polarization may shed light on the deep understanding in nuclear physics, astrophysics, high-energy physics etc.

Attosecond light beams can currently be generated by various methods, including high harmonic generation \cite{hentschel2001attosecond, Paul2001Observation,sansone2011high}, in free electron lasers (FEL) \cite{Saldin2004TW, Zholents2005Atto,xu2023ultracompact}, and as a result of interaction of ultra-intense lasers with plasmas.
The first two methods can generate photons with energies in the X-ray band or lower, whereas $\gamma-$rays with attosecond duration can be produced by the last method \cite{gu2018brilliant,lecz2019attosecond,zhang2021brilliant}.
Here, the peak intensity of the laser, up to $10^{22}\sim10^{23}$~W/cm$^2$, plays a crucial role.
These intensity requirements are available today at 10-PW scale laser facilities \cite{yanovsky2008ultra,pirozhkov2017approaching,guo2018improvement,tiwari2019beam,yoon2019achieving,yoon2021realization}.
Attosecond $\gamma-$ray generation, employing schemes that involve two lasers, needs to take care of the spatiotemporal synchronization issues \cite{li2017ultra,zhang2022generation}.  Some schemes also require a very cleverly structured laser, even for intensity of $\sim10^{21}$ W/cm$^2$, as in the case of a circularly polarized Laguerre-Gaussian laser \cite{zhu2018bright}. 
Furthermore, the angular divergence of these existing sources is typically on the order of several to tens of degrees, due to the deflection by the transverse laser field \cite{zhu2019collimated, Hu2021Atto}, and the polarization of the emitted photons is overlooked.
The beam-plasma interaction scheme, which was once used in wakefield acceleration experiments \cite{Kumar2010Self-modulation, Vieira2012Transverse}, has recently been proposed to generate collimated brilliant $\gamma-$rays \cite{benedetti2018giant, Sampath2021Extremely}, but the pulse duration and polarization has been ignored. 
Thus, the realization of a collimated high-brilliance polarized attosecond $\gamma-$ray source is still a great challenge.

 \begin{figure}[t]
	\centering
	\includegraphics[width=\linewidth]{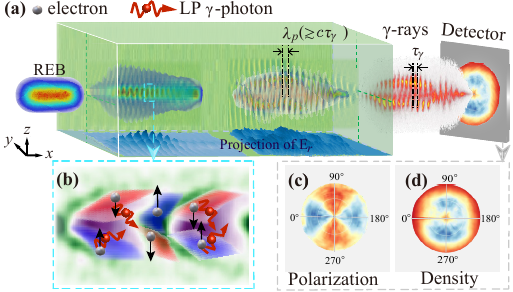}
	\caption{(a) The scenario of the REB interacting with a solid-density plasma. $\lambda_p$ indicates plasma wavelength and $\tau_\gamma$ indicates duration of $\gamma-$photon slice. (b) Details of modulation on the $\hat{x}$$-$$\hat{z}$ plane ($y=0$) and NCS process. Background is the plasma and electric field; black arrows represent forces on the electrons in the electric fields. (c) Schematic of polarization distribution, and (d) photon density, on $\theta-\varphi$ plane, where $\theta$ and $\varphi$ are polar and azimuthal angles, respectively.}
	\label{fig1}
\end{figure}
	
\begin{figure}[t]
	\includegraphics[width=\linewidth]{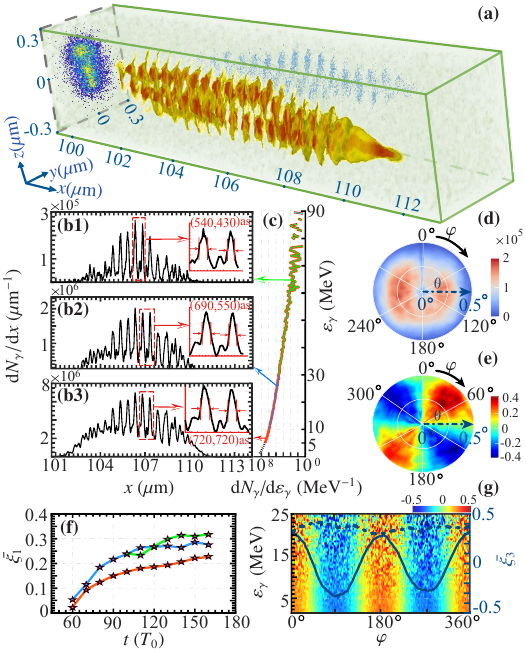}
	\caption{(a) 3D scenario of $\gamma-$photon pulse in the plasma at interaction time $t=100T_0$ and its projection onto $\hat{y}-\hat{z}$ plane and $\hat{x}-$direction.  (b1) Photon density $\mathrm{d}N_\gamma/\mathrm{d}x~(\mu\mathrm{m}^{-1})$ along the $\hat{x}-$direction corresponding to photon energies $\varepsilon_\gamma\geq30$~MeV, (b2) $\varepsilon_\gamma\in[10,30)$~MeV, and (b3) $\varepsilon_\gamma\geq5$~MeV. (c) Energy spectra of all photons at  $t=100T_0$.  (d) Angular density distribution $\mathrm{log_{10}}(\mathrm{d}^2N_\gamma/\mathrm{d}\theta \mathrm{d}\varphi)~(\mathrm{deg}^{-2})$ and (e) $\bar{\xi_1}$ of all photons with respect to polar angle $\theta$ and azimuthal angle $\varphi$. (f)  Average linear polarization $\bar{\xi_1}$ of attosecond $\gamma-$slice vs $t$, where the statistics are based on right peaks in insets of (b1)-(b3), respectively. Colored lines represent $\bar{\xi_1}$ of different energy ranges in (c). (g) Photon polarization $\bar{\xi_3}$ vs $\varepsilon_\gamma$ and  $\varphi$. $\bar{\xi_3}$ (solid line) and $\bar{\xi}=\sqrt{\overline{\xi}_1^2+\overline{\xi}_3^2}$ (dashed line) vs $\varphi$ with $\varepsilon_\gamma\geq5$~MeV.}
	\label{fig2}
        \vspace{-0.18cm} 
\end{figure}

\begin{figure}[!t]
	\includegraphics[width=\linewidth]{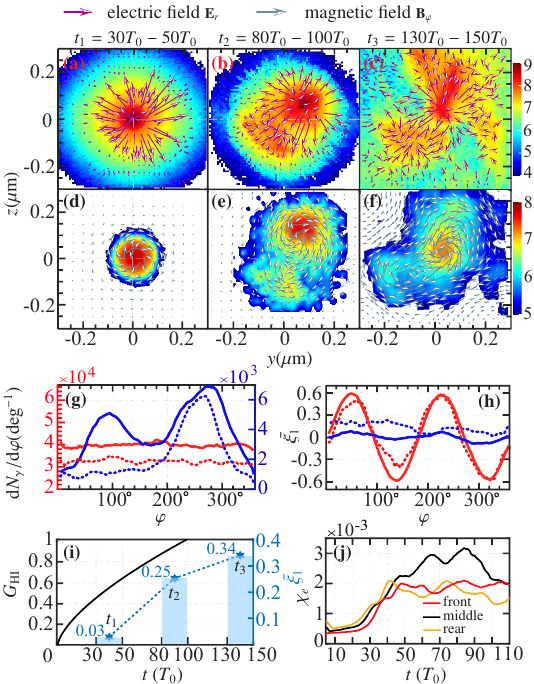}
	\caption{(a)-(c): Average number density $\log_{10}(\mathrm{d}^2N_e/\mathrm{d}y\mathrm{d}z)$ $(\mu\mathrm{m}^{-2})$ of a representative electron slice corresponding to the time intervals $t_1$, $t_2$ and $t_3$, respectively. This slice is selected from electrons situated at $x\in[36.625,36.935]$ at $t=30T_0$. Purple-arrows represent the electric field. (d)-(f): Average number density $\log_{10}(\mathrm{d}^2N_\gamma/\mathrm{d}y\mathrm{d}z)$ $(\mu\mathrm{m}^{-2})$ of photons radiated by the representative electron slice during $t_1$, $t_2$ and $t_3$, respectively. Gray-arrows represent  magnetic field.  (g) and (h): Angle-resolved density $\mathrm{d}N_{\gamma}/\mathrm{d}\varphi$ and $\bar{\xi_1}$ of photons emitted in $t_1$ (red lines) and $t_2$ (blue lines) vs $\varphi$, respectively. Solid lines and dashed-lines denote the results for photons radiated by the entire electron beam and by the slices depicted in (a) and (b). (i) Instability growth rate $G_{\mathrm{HI}}$ vs $t$, and $\bar{\xi_1}$ for the generated photons collected from the right peak of the inset in Fig.~\ref{fig2}(b3), during $t_1$, $t_2$ and $t_3$, respectively.  (j) Temporal evolution of QED parameter $\chi_e$ for three clumps of sampled electrons at the front, middle and rear of the REB. The sampled electrons are located at  $[108.688,108.708]$, $[106.734,106.754]$ and $[104.138,104.158]$, respectively, at $t=100T_0$. }
	\label{fig3}
	\vspace{-0.18cm} 
\end{figure}

\begin{figure}[tbp]
\includegraphics[width=\linewidth]{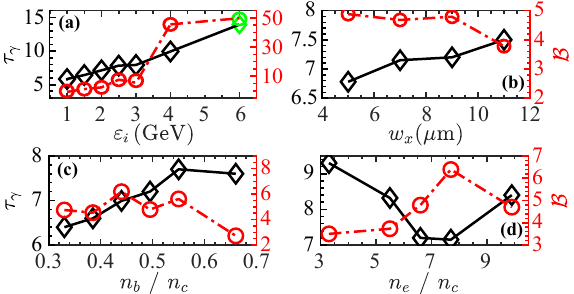}
	\caption{(a)-(d): Duration $\tau_\gamma$  (normalized by 100 as) and peak brilliance $\mathcal{B}$ [normalized by $10^{23}~\mathrm{photons/(s\cdot mm^{2}\cdot mrad^{2}\cdot0.1\%BW})$] of the generated $\gamma-$photon slices, vs the initial energy $\varepsilon_i$, length $w_x$ and density $n_b$ of the REB, and the plasma density $n_e$, respectively.  The green markers in (a) are obtained with all photon energies $\varepsilon_\gamma\geq10$~MeV. All other results are for 
all photon energies $\varepsilon_\gamma\geq5$ MeV.}
	\label{fig4}
\end{figure}

 In this Letter, we put forward a novel method to generate ultra-collimated high-brilliance polarized attosecond $\gamma-$rays via the direct interaction of an electron beam from laser-wakefield acceleration with a plasma; see the interaction scenario in Fig.~\ref{fig1}(a). 
 A relativistic electron beam (REB) is incident into a solid-density plasma of ionized polystyrene,  and excites periodic wakefields.
Self-focusing effect, induced by the plasma wakefield and self-field of the REB, modulates the REB into clusters, via the self-modulation instability (SMI), and further into attosecond slices, via the hosing instability (HI).
Meanwhile, these slices can emit linearly polarized high-energy $\gamma-$photons via nonlinear Compton scattering (NCS) while interacting with the wakefield; see Fig.~\ref{fig1}(b).
In principle , polarization orientation of the emitted $\gamma-$photons should be radially symmetric and therefore yields no net polarization; see Fig.~\ref{fig1}(c). 
However,  as the REB propagates in the plasma, HI dominates the transverse modulation of the REB, and breaks the transverse homogeneity of the REB as well as the distribution of the emitted $\gamma-$photons; see Fig.~\ref{fig1}(d).
For the given parameters, the angle-averaged degree of linear polarization of the $\gamma-$photon slice can reach about $\sim16.5\%$ (or 45\% within azimuthal angle $\varphi\in[40^\circ,80^\circ]$), with an angular divergence of only $9~\mathrm{mrad}$,  brilliance of  
 above $10^{23} ~\mathrm{photons/(s\cdot mm^{2}\cdot mrad^{2}\cdot0.1\%BW})$, and pulse duration of about hundreds of attoseconds; see more details in Figs.~\ref{fig2} and \ref{fig3}.
The proposed scheme is stable with respect to the REB and plasma parameters (see more details in Fig.~\ref{fig4}) and easy-to-deploy and, therefore, can be readily realized  on most moderately intense laser facilities. 
In addition, in the case of weak NCS and instabilities, our proposed method can also obtain high-quality electron microbunches which may serve as an alternative source for prebunched FELs \cite{wang2019angular}; see more details in \cite{Supplemental-Materials}. 

To simulate the spin- and polarization-resolved quantum electrodynamical (QED) processes, we use our QED particle-in-cell (PIC) code SLIPs \cite{wan2023Simulations,xuekun2023Generation} to simulate the beam-plasma interaction process.
We employ the invariant field parameter $a_0\equiv eE_0/(m_ec \omega)$, and the nonlinear quantum parameter $\chi_e\equiv(e\hslash/m_e^3c^4)\sqrt{-(F_{\mu\nu}p^\nu)^2}$.
The radiation probability based on the local constant field approximation \cite{ritus1985quantum,katkov1998electromagnetic,Piazza2019Improved}  is employed to calculate the electron radiation.
Here, $m_e$, $-e$ and $p^\nu$  denote the mass, charge and four-momentum of the electron, respectively, $F_{\mu \nu}$, $E_0$ and $\omega$ are the field tensor, amplitude and frequency of the external fields, respectively, $c$ is the light speed in vacuum, and $\hslash$ the reduced Planck constant.

In all simulations,  we use a simulation box with a spatial size of 15$\lambda_0(x)\times 0.6\lambda_0(y)\times 0.6\lambda_0(z)$,  divided into $960\times30\times30$ cells, and moving along $+\hat{x}$ direction, where $\lambda_0=1~\mu$m. 
Such spatial discretization can resolve the plasma skin-depth $k_p^{-1}$, where $k_p=\omega_p/c$, $\omega_p=\sqrt{n_e e^2 / (\epsilon_0 m_e)}$ is the plasma frequency, $\epsilon_0$ is vacuum permittivity, and $\lambda_p=2\pi c/\omega_p$ is the plasma wavelength.
We employ a REB with initial energy of 2~GeV, a flat-top longitudinal distribution with beam length of $w_x=9~\mathrm{\mu m}$, Gaussian transverse distribution with full-width-at-half-maximum of $0.2~\mu$m, and peak density of $n_b=0.45n_c$, where $n_c=(\omega_0^2m_e\epsilon_0)/e^2\approx1.1\times10^{21}$~cm$^{-3}$ is the critical plasma density with respect to the wavelength of $\lambda_0$,  $\omega_0=2 \pi c/\lambda_0$, and $T_0=\lambda_0/c$.
Such REB can be obtained via laser-wakefield acceleration by employing a moderately intense laser with a peak intensity of $10^{18}\sim10^{19}$~W/cm$^2$ \cite{lu2007generating,leemans2014multi}. Alternatively, direct laser acceleration can also be employed to generate the REB \cite{Babjak2024Direct}. 
The solid-density plasma target is assumed to be a fully ionized polystyrene foam with electron density $n_e=6n_c$.
The numbers of macro-particles per cell for background electrons, protons, carbon ions and beam electrons are 10, 3, 3 and 4, respectively.
Absorption boundaries are used in the transverse directions ($\hat{y}$ and $\hat{z}$) for both particles and fields.

Sample results are shown in Fig.~\ref{fig2}. 
Three dimensional (3D) scenario of the generated $\gamma-$photon pulse is presented in Fig.~\ref{fig2}(a) in which up to 8 attosecond slices are periodically distributed longitudinally and, due to HI, transversely aligned with the $\hat{z}$ direction; also see \cite{Supplemental-Materials}.
To meet the requirements of different applications, e.g., in photonuclear interactions, we have performed filtering on the photon energies of $\varepsilon_\gamma\geq30$~MeV and $\varepsilon_\gamma\in[10,30)$~MeV; see Figs.~\ref{fig2}(b1) and (b2). 
We also present the density distribution of all photons with  $\varepsilon_\gamma \geq 5~\mathrm{MeV}$; see Fig.~\ref{fig2}(b3). 
 As $\varepsilon_\gamma$ increases, the duration of the $\gamma-$photon slice  decreases. 
For $\varepsilon_\gamma\geq30$~MeV and $\varepsilon_\gamma\in[10,30)$~MeV, the minimum slice duration is about 430 as and 550 as, which could be suitable for the exploration of the quasi-deuteron excitation \cite{Young1998Evaluation,quasi-deuteron-international2000iaea} and giant dipole resonance (GDR) \cite{Chakrabarty2016GDR}, respectively. 
Slices of $\varepsilon_\gamma \geq 5~\mathrm{MeV}$, with a duration of the order of $720~\textrm{as}$ could be used in photoelectric disintegration studies \cite{Katsuma2014Photoelectric}.
The corresponding photon numbers [peak brilliance in units of $~\mathrm{photons/(s\cdot mm^{2}\cdot mrad^{2}\cdot0.1\%BW})$] can reach about $4.27\times 10^5 ~(1.1 \times 10^{23}$) at 30 MeV, $6.13\times10^6$ ($3.1\times10^{23}$) at 15 MeV, and $1.72\times10^7$ ($4.8\times10^{23}$) at 5 MeV, respectively; see Fig.~\ref{fig2}(c) and right peaks in insets of Figs.~\ref{fig2}(b1)-(b3).

Before $60T_0$, density of the REB in the transverse plane is uniform due to weak beam instabilities (SMI and HI), and so is angular distribution of the generated $\gamma-$photons.
The emitted photons are linearly polarized along the radial direction; see Fig.~\ref{fig2}(e).
For instance, as $t\lesssim 60T_0$, the angle-averaged polarization is below 3\%; see Fig.~\ref{fig2}(f).
After $60T_0$, transverse homogeneities of the REB and generated $\gamma-$photons are broken due to growing beam instabilities, as shown in Fig.~\ref{fig2}(d).
Angular asymmetry in the photon density distribution induces an increase in linear polarization of the $\gamma-$photon slices; see the temporal evolution of $\overline{\xi}_1$ in Fig.~\ref{fig2}(f), where $\xi_1$ and $\xi_3$ denote linear polarization along $45^\circ$ and  $0^\circ$ with respect to $\hat
{y}$ direction, respectively; see \cite {Supplemental-Materials}.
The final angle-averaged $\overline{\xi}_1$ can reach about $0.24$, $0.27$ and $0.32$ for photon energies over $30$ MeV, $15$ MeV and $5$ MeV, respectively; see Fig.~\ref{fig2}(f).
In addition to be used as a polarized attosecond $\gamma-$ray source,  these photons can also be used as a  byproduct to  measure the development level of the beam instability. 
The angular distribution of $\overline{\xi}_3$ shows a shift of $\pi/4$ with respect to $\overline{\xi}_1$, and stays almost uniform over the energy range $\varepsilon_\gamma\in[1,25]$~MeV; see Figs.~\ref{fig2}(e) and (g). 
Whereas the angle-resolved total linear polarization $\bar{\xi}=\sqrt{\overline{\xi}_1^2+\overline{\xi}_3^2}$ is $\sim38\%$ over the whole range of the azimuthal angle $\varphi$; see Fig.~\ref{fig2}(g).

Next, we analyze the generation of polarized attosecond $\gamma$-ray pulse train in detail. 
Depending on the growth of SMI and HI, we divide the whole interaction process into 3 stages. Stage I: REB slicing via the SMI and HI (before $60T_0$); stage II: stable radiation of the REB slice accompanied by the fast-growing SMI and HI ($60T_0-100T_0$);  and stage III: distortion of the REB slices due to strong HI (after $100T_0$).

    In stage I, a REB enters a solid-density plasma and plasma bubbles form with density perturbations $\delta n_e(x,t)\propto\delta n_{e,0}~\mathrm{sin}k_p\zeta$, in which $\delta n_e(x,t)\equiv n^{\prime}_e-n_e$ is the local density perturbation with amplitude $\delta n_{e,0}$, $\zeta=x - ct$ is the longitudinally co-moving coordinate, and $n^{\prime}_e$ is perturbed electron density \cite{Downer2018iagnostics,Tajima1979Laser}.
In this process, the plasma electrons are pushed outward, while the ions are left immobile. 
Plasma wakefield is excited and coupled with the REB self-field; see the transverse electric field distribution in Fig.~\ref{fig1}(b).
The longitudinal electric field component of the wakefield can be estimated as $E_\parallel\simeq k_p\zeta/2E_{0}\simeq\frac{k_0\zeta}{2}E_{0}\sqrt{n_e/n_c}$  \cite{Leemans2009Physics}, approximately $(10^{17}\times \zeta) ~\mathrm{V/m}$. 
Similarly, the transverse electric field is given by $E_\perp\simeq k_p r/4E_{0}\simeq \frac{k_0 r}{4}E_{0}\sqrt{n_e/n_c}$ , also of the same order. 
These are consistent with our simulations.
Here, $k_0=2\pi/\lambda_0$, and $r=\sqrt{y^2 + z^2}$ is the transverse coordinate.
Due to the relativistic effects, motion of the REB is dominated by the transverse field components \cite{Kumar2010Self-modulation}, as also can be inferred from the evolution process of the REB in the transverse plane depicted in Figs.~\ref{fig3}(a)-(c), with more details in \cite{Supplemental-Materials}.
We also confirm that, for the parameters used here,  the electrons are modulated primarily by the electric field, rather than by the electromagnetic coupling field, until the growth rates of beam instabilities reach approximately 1; see \cite{Supplemental-Materials}.
The REB experiences alternate focusing and defocusing fields in the transverse direction, which will induce periodic modulation via the SMI \cite{Schroeder2011Growth, Pukhov2011Phase}, thus turning the REB into high-density clusters, each with a scale length comparable to the plasma wavelength $\lambda_p\propto n_e^{-1/2}$ ; see Fig.~\ref{fig1}(a).
In this stage, SMI and HI are relatively weak, and the transverse misplacement of REB is negligible; see Fig.~\ref{fig3}(a) for  density distribution of the REB.
Therefore, the azimuthal dependence of the transverse electric and magnetic fields is uniform; see Figs.~\ref{fig3}(a) and (d). 
The emitted $\gamma-$photons via the NCS of electrons, which follow the REB, are uniformly distributed across the transverse plane; see Figs.~\ref{fig3}(a), (d) and (g).
Besides, they exhibit centrosymmetric radial linear polarization at azimuthal angle. 
This means that polarized $\gamma-$rays can be obtained by selecting specific azimuthal angles; see Fig.~\ref{fig3}(h). However, upon averaging the photon polarization over the entire angular space, the linear polarization is negligible; see Fig.~\ref{fig3}(i).  

In stage II, due to electrostatic coupling of the transverse beam displacements with respect to the sheath electrons of the wakefield, the centroid of the REB oscillates around the propagation axis, i.e., HI begins to dominate the transverse beam dynamics \cite{Whittum1991Electron}.
The parameters we utilize satisfy the requirements $k_p\sigma_x\gg1, k_p\sigma_r\ll1$, and $n_b/n_e\ll1$, which are crucial for the development of SMI and HI. Consequently, the instabilities growth in our simulation can be described by $e^G$, with the growth rate $G$. 
For HI, $G=G_\mathrm{HI}\approx\frac{3^{3/2}}{2^{7/3}}(\frac{1}{\gamma_e}\frac{n_b}{n_e} k_p^3 \zeta x^2)^{1/3}$ \cite{Schroeder2012Coupled, Schroeder2011Growth}, and for SMI,  $G=G_\mathrm{SMI} \approx 2^{1/3} G_\mathrm{HI}$ \cite{Schroeder2012Coupled, Vieira2012Transverse}, respectively, where $\sigma_x$ and $\sigma_r$ are the rms length and radius of the  REB, respectively, $\gamma_e$ is Lorentz factor.
As mentioned previously, in stage I, due to the growth of SMI, the REB is modulated into clusters.
However, as the growth rates of SMI and HI are comparable, these high-density clusters get subsequently transformed into  staggered slices (owing to increased strength of the HI) with a smaller scale length than before, which is about $\lambda_p/2$, i.e., into attosecond slices for $n_e\gtrsim(\lambda_0^2/\lambda_p^2) n_c$ and $\lambda_p \lesssim 0.6~\mathrm{\mu m}$. 
See more details of the modulation process in \cite{Supplemental-Materials}.
As a result, during the modulation,  NCS $\gamma$-photons emitted by the REB exhibit a periodic structure with a period on the attosecond scale as well; see Figs.~\ref{fig1}(a) and \ref{fig2}(a).
For instance, with $n_e\simeq6n_c$, the pulse duration of each $\gamma$-slice can reach about $720~\mathrm{as}$.
Note that only 0.016\% of the plasma electrons are accelerated with a maximum energy of $\sim 0.82$ MeV, and the impact of radiation from the plasma electrons on our $\gamma-$ray yield is negligible.

Meanwhile, with the growth of the HI, displacement of the centroid of each slice in the transverse plane reaches a measurable magnitude, and axial symmetry of the electromagnetic field is broken simultaneously; see Figs.~\ref{fig3}(b) and (e).
Not only are the $\gamma$-photon slices emitted in this stage are in the attosecond scale, due to the axial symmetry breaking in the distribution of density and linear polarization, but also the net polarization can reach about $0.25$; see Figs.~\ref{fig3}(g)-(i). 
However, in stage III, owing to the fast growth of SMI and HI, the transverse beam profile becomes severely irregular and can no longer sustain the feature of attosecond slices, which, in turn, can not excite an effective wakefield; see Figs.~\ref{fig3}(c) and (f). 
As a result, NCS is suppressed due to much smaller $\chi_e$, but acquires a higher degree of linear polarization due to the high asymmetry of the transverse field; see Figs.~\ref{fig3}(i) and (j).
Further propagation of the REB in the plasma may increase the duration of the $\gamma-$photon slices, therefore, one should terminate the plasma to obtain a high-quality attosecond $\gamma-$pulse train.

To demonstrate experimental feasibility, we also study the impact of the initial density $n_b$, energy $\varepsilon_i$ and beam length $w_x$ of the REB, as well as the plasma density $n_e$, on the duration $\tau_\gamma$ and brilliance $\mathcal{B}$ of the $\gamma-$photon slices; see Fig.~\ref{fig4}.
For higher electron energies $\varepsilon_i$, on the one hand, the modulation process will be slowed down as $G_{\mathrm{HI}}\propto\gamma_e^{-1/3}$, and the radiation duration will increase before the REB is modulated into slices, i.e., more photons will be located outside the slice, thereby increasing the duration of each $\gamma-$photon slice.
An electron beam possessing an ultra-high energy, such as 6 GeV [see green markers in Fig.~\ref{fig4}(a)],  emits a significant number of low-energy photons during stage I. Thus, it is unable to produce short pulses in the low-energy range but can do so in the high-energy range. 
On the other hand, the radiation probability ($\propto\varepsilon_i$) will increase for electrons, which will enhance the photon emission to acquire higher brilliance; see Fig.~\ref{fig4}(a). 
Because the growth rate $G_{\mathrm{HI}}$ of HI does not explicitly depend on $w_x$, which therefore does not contribute to the growth of HI, as long as $k_p \sigma_x \gg 1$ (in our cases, $\sigma_x = w_x$). 
Under the influence of SMI, modulation of the REB occurs downstream first, and expands to the entire beam later.
Hence, keeping other parameters the same, a beam with a larger $w_x$ will require a longer modulation process, which finally will increase the duration of $\gamma$-photon slices.
Besides, prolonged modulation leads to an increase in the angular divergence of both electrons and emitted photons and decreases brilliance of the source; see Fig.~\ref{fig4}(b).

When employing a REB with a much higher density $n_b$, the photon yield before electron slicing will be much higher.
More photons will occupy the gaps between the formed $\gamma$-photon slices, thus increasing the duration of each $\gamma$-photon slice. 
As $n_b$ increases, $G_{\mathrm{HI}}\propto(n_b/n_e)^{1/3}$ increases as well, which reduces the stable radiation significantly before the disruption of the REB slices. 
As a result, though the electron density involved in the emission is higher, the increment of the radiation brilliance is negligible. 
On the contrary, with a higher density $n_b$, the radiation brilliance generally tends to decrease; see Fig.~\ref{fig4}(c).
The plasma density $n_e$ is another key factor for the radiation.
With larger $n_e$, $G_{\mathrm{HI}}$ will be smaller. 
This leads to a longer radiation time for the REB, either before or after the slice formation, and finally yields a higher brilliance for the $\gamma-$photon slices.
Meanwhile, the modulation scale length, i.e.,  thickness of the electron slice, which is less than the plasma wavelength $\lambda_p \propto n_e^{-1/2}$, will decrease, and so will duration of the photon slice $\tau_\gamma$.
For much higher $n_e$, e.g., $n_e=9n_c$, the REB is unable to excite periodic large-amplitude wakefields due to severe beam instabilities.
Therefore, the REB cannot be efficiently modulated, and the duration of the emitted photon slices increases (contradicting the decrease of $\lambda_p$).
Besides, with a stronger field excited in the plasma, oscillation amplitude of an electron will be larger, thereby inducing a larger emission angle. 
Thus, the photon angular divergence increases, and the brilliance decreases; see Fig.~\ref{fig4}(d). 
It is worth mentioning that polarization of the attosecond $\gamma$-ray beam depends on the strength of the instability, which exhibits weak dependence on the above parameter ranges.
Attosecond $\gamma-$rays maintain certain angle-averaged and angle-resolved degrees of polarization.
Despite its dependence on the initial energy, length, density of REB and plasma density, this scheme exhibits a good degree of robustness over a wide range of parameter values.
In addition, the generated series of electron slices, typically with a thickness of tens of nanometers and an energy spread of $\sim0.23\%$, may serve as a potential source for prebunched FELs; see \cite{Supplemental-Materials}.

In conclusion, we have put forward a novel and easy-to-deploy scheme to generate ultra-collimated high-brilliance polarized attosecond $\gamma-$rays, via the interaction of a laser-wakefield-acceleration electron beam with a solid-density plasma.
Such high-quality beams have significant applications in nuclear physics, astrophysics, high-energy physics, etc. Examples include their potential to unveil new physics beyond the Standard Model through ultrahigh-precision tracing of intricate reactions, or to serve as a source of polarized attosecond positron and electron beams.  
The degree of photon polarization, caused by the beam instability, provides a new perspective for measuring or estimating the evolution of instability in a plasma.
Additionally, the mechanism of the beam modulation period related to plasma density renders our scheme promising for generating shorter pulses, such as zeptosecond pulses.

 ~~~~~~~~~~

\emph{Acknowledgments:} This work is supported by the National Natural Science Foundation of China (Grants No. U2267204, No. 12275209), the Foundation of Science and Technology on Plasma Physics Laboratory (No. JCKYS2021212008), the Shaanxi Fundamental Science Research Project for Mathematics and Physics (Grant No. 22JSY014), and the Fundamental Research Funds for Central Universities (No. xzy012023046). Lei-Feng Cao is financially supported by  Outstanding Talents Training Fund in Shenzhen (No. 202101) and National Key R\&D Program Project (No. 2022YFA1603302).

\bibliography{atto-polar}

\begin{thebibliography}{63}%
\makeatletter
\providecommand \@ifxundefined [1]{%
 \@ifx{#1\undefined}
}%
\providecommand \@ifnum [1]{%
 \ifnum #1\expandafter \@firstoftwo
 \else \expandafter \@secondoftwo
 \fi
}%
\providecommand \@ifx [1]{%
 \ifx #1\expandafter \@firstoftwo
 \else \expandafter \@secondoftwo
 \fi
}%
\providecommand \natexlab [1]{#1}%
\providecommand \enquote  [1]{``#1''}%
\providecommand \bibnamefont  [1]{#1}%
\providecommand \bibfnamefont [1]{#1}%
\providecommand \citenamefont [1]{#1}%
\providecommand \href@noop [0]{\@secondoftwo}%
\providecommand \href [0]{\begingroup \@sanitize@url \@href}%
\providecommand \@href[1]{\@@startlink{#1}\@@href}%
\providecommand \@@href[1]{\endgroup#1\@@endlink}%
\providecommand \@sanitize@url [0]{\catcode `\\12\catcode `\$12\catcode
  `\&12\catcode `\#12\catcode `\^12\catcode `\_12\catcode `\%12\relax}%
\providecommand \@@startlink[1]{}%
\providecommand \@@endlink[0]{}%
\providecommand \url  [0]{\begingroup\@sanitize@url \@url }%
\providecommand \@url [1]{\endgroup\@href {#1}{\urlprefix }}%
\providecommand \urlprefix  [0]{URL }%
\providecommand \Eprint [0]{\href }%
\providecommand \doibase [0]{https://doi.org/}%
\providecommand \selectlanguage [0]{\@gobble}%
\providecommand \bibinfo  [0]{\@secondoftwo}%
\providecommand \bibfield  [0]{\@secondoftwo}%
\providecommand \translation [1]{[#1]}%
\providecommand \BibitemOpen [0]{}%
\providecommand \bibitemStop [0]{}%
\providecommand \bibitemNoStop [0]{.\EOS\space}%
\providecommand \EOS [0]{\spacefactor3000\relax}%
\providecommand \BibitemShut  [1]{\csname bibitem#1\endcsname}%
\let\auto@bib@innerbib\@empty
\bibitem [{\citenamefont {Villeneuve}(2018)}]{villeneuve2018Atto}%
  \BibitemOpen
  \bibfield  {author} {\bibinfo {author} {\bibfnamefont {D.~M.}\ \bibnamefont
  {Villeneuve}},\ }\bibfield  {title} {\bibinfo {title} {Attosecond science},\
  }\href {https://doi.org/10.1080/00107514.2017.1407093} {\bibfield  {journal}
  {\bibinfo  {journal} {Contemp. Phys.}\ }\textbf {\bibinfo {volume} {59}},\
  \bibinfo {pages} {47} (\bibinfo {year} {2018})}\BibitemShut {NoStop}%
\bibitem [{\citenamefont {Zewail}(2000)}]{zewail2000femtochemistry}%
  \BibitemOpen
  \bibfield  {author} {\bibinfo {author} {\bibfnamefont {A.~H.}\ \bibnamefont
  {Zewail}},\ }\bibfield  {title} {\bibinfo {title} {{Femtochemistry:
  Atomic-scale dynamics of the chemical bond}},\ }\href
  {https://doi.org/https://doi.org/10.1021/jp001460h} {\bibfield  {journal}
  {\bibinfo  {journal} {J. Phys. Chem. A}\ }\textbf {\bibinfo {volume} {104}},\
  \bibinfo {pages} {5660} (\bibinfo {year} {2000})}\BibitemShut {NoStop}%
\bibitem [{\citenamefont {Corkum}\ and\ \citenamefont
  {Krausz}(2007)}]{corkum2007attosecond}%
  \BibitemOpen
  \bibfield  {author} {\bibinfo {author} {\bibfnamefont {P.~B.}\ \bibnamefont
  {Corkum}}\ and\ \bibinfo {author} {\bibfnamefont {F.}~\bibnamefont
  {Krausz}},\ }\bibfield  {title} {\bibinfo {title} {Attosecond science},\
  }\href {https://doi.org/https://doi.org/10.1038/nphys620} {\bibfield
  {journal} {\bibinfo  {journal} {Nat. phys.}\ }\textbf {\bibinfo {volume}
  {3}},\ \bibinfo {pages} {381} (\bibinfo {year} {2007})}\BibitemShut {NoStop}%
\bibitem [{\citenamefont {Kling}\ and\ \citenamefont
  {Vrakking}(2008)}]{Kling2008Atto}%
  \BibitemOpen
  \bibfield  {author} {\bibinfo {author} {\bibfnamefont {M.~F.}\ \bibnamefont
  {Kling}}\ and\ \bibinfo {author} {\bibfnamefont {M.~J.}\ \bibnamefont
  {Vrakking}},\ }\bibfield  {title} {\bibinfo {title} {{Attosecond Electron
  Dynamics}},\ }\href
  {https://doi.org/https://doi.org/10.1146/annurev.physchem.59.032607.093532}
  {\bibfield  {journal} {\bibinfo  {journal} {Annu. Rev. Phys. Chem.}\ }\textbf
  {\bibinfo {volume} {59}},\ \bibinfo {pages} {463} (\bibinfo {year}
  {2008})}\BibitemShut {NoStop}%
\bibitem [{\citenamefont {Drescher}\ \emph {et~al.}(2002)\citenamefont
  {Drescher}, \citenamefont {Hentschel}, \citenamefont {Kienberger},
  \citenamefont {Uiberacker}, \citenamefont {Yakovlev}, \citenamefont
  {Scrinzi}, \citenamefont {Westerwalbesloh}, \citenamefont {Kleineberg},
  \citenamefont {Heinzmann},\ and\ \citenamefont {Krausz}}]{drescher2002time}%
  \BibitemOpen
  \bibfield  {author} {\bibinfo {author} {\bibfnamefont {M.}~\bibnamefont
  {Drescher}}, \bibinfo {author} {\bibfnamefont {M.}~\bibnamefont {Hentschel}},
  \bibinfo {author} {\bibfnamefont {R.}~\bibnamefont {Kienberger}}, \bibinfo
  {author} {\bibfnamefont {M.}~\bibnamefont {Uiberacker}}, \bibinfo {author}
  {\bibfnamefont {V.}~\bibnamefont {Yakovlev}}, \bibinfo {author}
  {\bibfnamefont {A.}~\bibnamefont {Scrinzi}}, \bibinfo {author} {\bibfnamefont
  {T.}~\bibnamefont {Westerwalbesloh}}, \bibinfo {author} {\bibfnamefont
  {U.}~\bibnamefont {Kleineberg}}, \bibinfo {author} {\bibfnamefont
  {U.}~\bibnamefont {Heinzmann}},\ and\ \bibinfo {author} {\bibfnamefont
  {F.}~\bibnamefont {Krausz}},\ }\bibfield  {title} {\bibinfo {title}
  {Time-resolved atomic inner-shell spectroscopy},\ }\href
  {https://doi.org/https://doi.org/10.1038/nature01143} {\bibfield  {journal}
  {\bibinfo  {journal} {Nature}\ }\textbf {\bibinfo {volume} {419}},\ \bibinfo
  {pages} {803} (\bibinfo {year} {2002})}\BibitemShut {NoStop}%
\bibitem [{\citenamefont {Baltuška}\ \emph {et~al.}(2003)\citenamefont
  {Baltuška}, \citenamefont {Udem}, \citenamefont {Uiberacker}, \citenamefont
  {Hentschel}, \citenamefont {Goulielmakis}, \citenamefont {Gohle},
  \citenamefont {Holzwarth}, \citenamefont {Yakovlev}, \citenamefont {Scrinzi},
  \citenamefont {Hänsch},\ and\ \citenamefont
  {Krausz}}]{baltuvska2003attosecond_control}%
  \BibitemOpen
  \bibfield  {author} {\bibinfo {author} {\bibfnamefont {A.}~\bibnamefont
  {Baltuška}}, \bibinfo {author} {\bibfnamefont {T.}~\bibnamefont {Udem}},
  \bibinfo {author} {\bibfnamefont {M.}~\bibnamefont {Uiberacker}}, \bibinfo
  {author} {\bibfnamefont {M.}~\bibnamefont {Hentschel}}, \bibinfo {author}
  {\bibfnamefont {E.}~\bibnamefont {Goulielmakis}}, \bibinfo {author}
  {\bibfnamefont {C.}~\bibnamefont {Gohle}}, \bibinfo {author} {\bibfnamefont
  {R.}~\bibnamefont {Holzwarth}}, \bibinfo {author} {\bibfnamefont {V.~S.}\
  \bibnamefont {Yakovlev}}, \bibinfo {author} {\bibfnamefont {A.}~\bibnamefont
  {Scrinzi}}, \bibinfo {author} {\bibfnamefont {T.~W.}\ \bibnamefont
  {Hänsch}},\ and\ \bibinfo {author} {\bibfnamefont {F.}~\bibnamefont
  {Krausz}},\ }\bibfield  {title} {\bibinfo {title} {Attosecond control of
  electronic processes by intense light fields},\ }\href
  {https://doi.org/https://doi.org/10.1038/nature01414} {\bibfield  {journal}
  {\bibinfo  {journal} {Nature}\ }\textbf {\bibinfo {volume} {421}},\ \bibinfo
  {pages} {611} (\bibinfo {year} {2003})}\BibitemShut {NoStop}%
\bibitem [{\citenamefont {Drescher}\ \emph {et~al.}(2001)\citenamefont
  {Drescher}, \citenamefont {Hentschel}, \citenamefont {Kienberger},
  \citenamefont {Tempea}, \citenamefont {Spielmann}, \citenamefont {Reider},
  \citenamefont {Corkum},\ and\ \citenamefont {Krausz}}]{Markus2001X-ray}%
  \BibitemOpen
  \bibfield  {author} {\bibinfo {author} {\bibfnamefont {M.}~\bibnamefont
  {Drescher}}, \bibinfo {author} {\bibfnamefont {M.}~\bibnamefont {Hentschel}},
  \bibinfo {author} {\bibfnamefont {R.}~\bibnamefont {Kienberger}}, \bibinfo
  {author} {\bibfnamefont {G.}~\bibnamefont {Tempea}}, \bibinfo {author}
  {\bibfnamefont {C.}~\bibnamefont {Spielmann}}, \bibinfo {author}
  {\bibfnamefont {G.~A.}\ \bibnamefont {Reider}}, \bibinfo {author}
  {\bibfnamefont {P.~B.}\ \bibnamefont {Corkum}},\ and\ \bibinfo {author}
  {\bibfnamefont {F.}~\bibnamefont {Krausz}},\ }\bibfield  {title} {\bibinfo
  {title} {{X-ray Pulses Approaching the Attosecond Frontier}},\ }\href
  {https://api.semanticscholar.org/CorpusID:31720358} {\bibfield  {journal}
  {\bibinfo  {journal} {Science}\ }\textbf {\bibinfo {volume} {291}},\ \bibinfo
  {pages} {1923 } (\bibinfo {year} {2001})}\BibitemShut {NoStop}%
\bibitem [{\citenamefont {Utsunomiya}\ \emph {et~al.}(2003)\citenamefont
  {Utsunomiya}, \citenamefont {Akimune}, \citenamefont {Goko}, \citenamefont
  {Ohta}, \citenamefont {Ueda}, \citenamefont {Yamagata}, \citenamefont
  {Yamasaki}, \citenamefont {Ohgaki}, \citenamefont {Toyokawa}, \citenamefont
  {Lui}, \citenamefont {Hayakawa}, \citenamefont {Shizuma}, \citenamefont
  {Khan},\ and\ \citenamefont {Goriely}}]{utsunomiya2003cross}%
  \BibitemOpen
  \bibfield  {author} {\bibinfo {author} {\bibfnamefont {H.}~\bibnamefont
  {Utsunomiya}}, \bibinfo {author} {\bibfnamefont {H.}~\bibnamefont {Akimune}},
  \bibinfo {author} {\bibfnamefont {S.}~\bibnamefont {Goko}}, \bibinfo {author}
  {\bibfnamefont {M.}~\bibnamefont {Ohta}}, \bibinfo {author} {\bibfnamefont
  {H.}~\bibnamefont {Ueda}}, \bibinfo {author} {\bibfnamefont {T.}~\bibnamefont
  {Yamagata}}, \bibinfo {author} {\bibfnamefont {K.}~\bibnamefont {Yamasaki}},
  \bibinfo {author} {\bibfnamefont {H.}~\bibnamefont {Ohgaki}}, \bibinfo
  {author} {\bibfnamefont {H.}~\bibnamefont {Toyokawa}}, \bibinfo {author}
  {\bibfnamefont {Y.-W.}\ \bibnamefont {Lui}}, \bibinfo {author} {\bibfnamefont
  {T.}~\bibnamefont {Hayakawa}}, \bibinfo {author} {\bibfnamefont
  {T.}~\bibnamefont {Shizuma}}, \bibinfo {author} {\bibfnamefont
  {E.}~\bibnamefont {Khan}},\ and\ \bibinfo {author} {\bibfnamefont
  {S.}~\bibnamefont {Goriely}},\ }\bibfield  {title} {\bibinfo {title} {{Cross
  section measurements of the
  ${}^{181}\mathrm{Ta}(\ensuremath{\gamma}{,n)}^{180}\mathrm{Ta}$ reaction near
  neutron threshold and the $p$-process nucleosynthesis}},\ }\href
  {https://doi.org/10.1103/PhysRevC.67.015807} {\bibfield  {journal} {\bibinfo
  {journal} {Phys. Rev. C}\ }\textbf {\bibinfo {volume} {67}},\ \bibinfo
  {pages} {015807} (\bibinfo {year} {2003})}\BibitemShut {NoStop}%
\bibitem [{\citenamefont {Xu}\ \emph {et~al.}(2007)\citenamefont {Xu},
  \citenamefont {Xu}, \citenamefont {Ma}, \citenamefont {Guo}, \citenamefont
  {Chen}, \citenamefont {Cai}, \citenamefont {Wang}, \citenamefont {Wang},
  \citenamefont {Lu},\ and\ \citenamefont {Shen}}]{xu2007new}%
  \BibitemOpen
  \bibfield  {author} {\bibinfo {author} {\bibfnamefont {Y.}~\bibnamefont
  {Xu}}, \bibinfo {author} {\bibfnamefont {W.}~\bibnamefont {Xu}}, \bibinfo
  {author} {\bibfnamefont {Y.}~\bibnamefont {Ma}}, \bibinfo {author}
  {\bibfnamefont {W.}~\bibnamefont {Guo}}, \bibinfo {author} {\bibfnamefont
  {J.}~\bibnamefont {Chen}}, \bibinfo {author} {\bibfnamefont {X.}~\bibnamefont
  {Cai}}, \bibinfo {author} {\bibfnamefont {H.}~\bibnamefont {Wang}}, \bibinfo
  {author} {\bibfnamefont {C.}~\bibnamefont {Wang}}, \bibinfo {author}
  {\bibfnamefont {G.}~\bibnamefont {Lu}},\ and\ \bibinfo {author}
  {\bibfnamefont {W.}~\bibnamefont {Shen}},\ }\bibfield  {title} {\bibinfo
  {title} {{A new study for $^{16}${O}($\gamma$, $\alpha$)$^{12}${C} at the
  energies of nuclear astrophysics interest: The inverse of key nucleosynthesis
  reaction $^{12}${C}($\alpha$, $\gamma$)$^{16}${O}}},\ }\href
  {https://doi.org/https://doi.org/10.1016/j.nima.2007.08.078} {\bibfield
  {journal} {\bibinfo  {journal} {Nucl. Instrum. Methods A}\ }\textbf {\bibinfo
  {volume} {581}},\ \bibinfo {pages} {866} (\bibinfo {year}
  {2007})}\BibitemShut {NoStop}%
\bibitem [{\citenamefont {Weller}\ \emph {et~al.}(1992)\citenamefont {Weller},
  \citenamefont {Langenbrunner}, \citenamefont {Chasteler}, \citenamefont
  {Tomusiak}, \citenamefont {Asai}, \citenamefont {Seyler},\ and\ \citenamefont
  {Lehman}}]{weller1992angular}%
  \BibitemOpen
  \bibfield  {author} {\bibinfo {author} {\bibfnamefont {H.}~\bibnamefont
  {Weller}}, \bibinfo {author} {\bibfnamefont {J.}~\bibnamefont
  {Langenbrunner}}, \bibinfo {author} {\bibfnamefont {R.}~\bibnamefont
  {Chasteler}}, \bibinfo {author} {\bibfnamefont {E.}~\bibnamefont {Tomusiak}},
  \bibinfo {author} {\bibfnamefont {J.}~\bibnamefont {Asai}}, \bibinfo {author}
  {\bibfnamefont {R.}~\bibnamefont {Seyler}},\ and\ \bibinfo {author}
  {\bibfnamefont {D.}~\bibnamefont {Lehman}},\ }\bibfield  {title} {\bibinfo
  {title} {{Angular distribution coefficients for ($\gamma$, \emph{X})
  reactions with linearly polarized photons}},\ }\href
  {https://doi.org/https://doi.org/10.1016/0092-640X(92)90024-C} {\bibfield
  {journal} {\bibinfo  {journal} {At. Data Nucl. Data Tables}\ }\textbf
  {\bibinfo {volume} {50}},\ \bibinfo {pages} {29} (\bibinfo {year}
  {1992})}\BibitemShut {NoStop}%
\bibitem [{Pho(2000)}]{Photonuclear-data}%
  \BibitemOpen
  \href
  {https://www.iaea.org/publications/6043/handbook-on-photonuclear-data-for-applications-cross-sections-and-spectra}
  {\emph {\bibinfo {title} {{Handbook on Photonuclear Data for Applications
  Cross-sections and Spectra}}}},\ \bibinfo {series} {TECDOC Series}\ No.\
  \bibinfo {number} {1178}\ (\bibinfo  {publisher} {INTERNATIONAL ATOMIC ENERGY
  AGENCY},\ \bibinfo {address} {Vienna},\ \bibinfo {year} {2000})\BibitemShut
  {NoStop}%
\bibitem [{\citenamefont {Kossov}(2002)}]{kossov2002approximation}%
  \BibitemOpen
  \bibfield  {author} {\bibinfo {author} {\bibfnamefont {M.}~\bibnamefont
  {Kossov}},\ }\bibfield  {title} {\bibinfo {title} {Approximation of
  photonuclear interaction cross-sections},\ }\href
  {https://doi.org/https://doi.org/10.1140/epja/i2002-10008-x} {\bibfield
  {journal} {\bibinfo  {journal} {Eur. Phys. J. A}\ }\textbf {\bibinfo {volume}
  {14}},\ \bibinfo {pages} {377} (\bibinfo {year} {2002})}\BibitemShut
  {NoStop}%
\bibitem [{\citenamefont {Speth}\ and\ \citenamefont {van~der
  Woude}(1981)}]{J.Speth_1981Giant}%
  \BibitemOpen
  \bibfield  {author} {\bibinfo {author} {\bibfnamefont {J.}~\bibnamefont
  {Speth}}\ and\ \bibinfo {author} {\bibfnamefont {A.}~\bibnamefont {van~der
  Woude}},\ }\bibfield  {title} {\bibinfo {title} {Giant resonances in
  nuclei},\ }\href {https://doi.org/10.1088/0034-4885/44/7/002} {\bibfield
  {journal} {\bibinfo  {journal} {Rep. Prog. Phys.}\ }\textbf {\bibinfo
  {volume} {44}},\ \bibinfo {pages} {719} (\bibinfo {year} {1981})}\BibitemShut
  {NoStop}%
\bibitem [{\citenamefont {Akbar}\ \emph {et~al.}(2017)\citenamefont {Akbar},
  \citenamefont {Roy}, \citenamefont {Park}, \citenamefont {Crede},
  \citenamefont {Anisovich}, \citenamefont {Denisenko}, \citenamefont {Klempt},
  \citenamefont {Nikonov}, \citenamefont {Sarantsev}, \citenamefont {Adhikari},
  \citenamefont {Adhikari}, \citenamefont {Amaryan}, \citenamefont
  {Anefalos~Pereira}, \citenamefont {Avakian}, \citenamefont {Ball},
  \citenamefont {Battaglieri}, \citenamefont {Batourine}, \citenamefont
  {Bedlinskiy}, \citenamefont {Boiarinov}, \citenamefont {Briscoe},
  \citenamefont {Brock}, \citenamefont {Brooks}, \citenamefont {Burkert},
  \citenamefont {Cao}, \citenamefont {Carlin}, \citenamefont {Carman},
  \citenamefont {Celentano}, \citenamefont {Charles}, \citenamefont {Chetry},
  \citenamefont {Ciullo}, \citenamefont {Clark}, \citenamefont {Cole},
  \citenamefont {Contalbrigo}, \citenamefont {Cortes}, \citenamefont
  {D'Angelo}, \citenamefont {Dashyan}, \citenamefont {De~Vita}, \citenamefont
  {De~Sanctis}, \citenamefont {Deur}, \citenamefont {Djalali}, \citenamefont
  {Dugger}, \citenamefont {Dupre}, \citenamefont {Egiyan}, \citenamefont
  {El~Fassi}, \citenamefont {Eugenio}, \citenamefont {Fedotov}, \citenamefont
  {Fersch}, \citenamefont {Filippi}, \citenamefont {Fradi}, \citenamefont
  {Gar\ifmmode~\mbox{\c{c}}\else \c{c}\fi{}on}, \citenamefont {Gevorgyan},
  \citenamefont {Giovanetti}, \citenamefont {Girod}, \citenamefont {Gleason},
  \citenamefont {Gohn}, \citenamefont {Golovatch}, \citenamefont {Gothe},
  \citenamefont {Griffioen}, \citenamefont {Guidal}, \citenamefont {Guo},
  \citenamefont {Hafidi}, \citenamefont {Hakobyan}, \citenamefont {Hanretty},
  \citenamefont {Harrison}, \citenamefont {Hattawy}, \citenamefont {Heddle},
  \citenamefont {Hicks}, \citenamefont {Hollis}, \citenamefont {Holtrop},
  \citenamefont {Ireland}, \citenamefont {Ishkhanov}, \citenamefont {Isupov},
  \citenamefont {Jenkins}, \citenamefont {Joosten}, \citenamefont {Keith},
  \citenamefont {Keller}, \citenamefont {Khachatryan}, \citenamefont
  {Khachatryan}, \citenamefont {Khandaker}, \citenamefont {Kim}, \citenamefont
  {Kim}, \citenamefont {Klein}, \citenamefont {Klein}, \citenamefont
  {Kubarovsky}, \citenamefont {Lanza}, \citenamefont {Livingston},
  \citenamefont {MacGregor}, \citenamefont {Markov}, \citenamefont {McKinnon},
  \citenamefont {Meekins}, \citenamefont {Mineeva}, \citenamefont {Mokeev},
  \citenamefont {Movsisyan}, \citenamefont {Munoz~Camacho}, \citenamefont
  {Nadel-Turonski}, \citenamefont {Niccolai}, \citenamefont {Osipenko},
  \citenamefont {Ostrovidov}, \citenamefont {Paolone}, \citenamefont
  {Paremuzyan}, \citenamefont {Park}, \citenamefont {Pasyuk}, \citenamefont
  {Phelps}, \citenamefont {Pogorelko}, \citenamefont {Price}, \citenamefont
  {Prok}, \citenamefont {Protopopescu}, \citenamefont {Raue}, \citenamefont
  {Ripani}, \citenamefont {Ritchie}, \citenamefont {Rizzo}, \citenamefont
  {Rosner}, \citenamefont {Sabati\'e}, \citenamefont {Salgado}, \citenamefont
  {Schumacher}, \citenamefont {Sharabian}, \citenamefont {Skorodumina},
  \citenamefont {Smith}, \citenamefont {Sober}, \citenamefont {Sokhan},
  \citenamefont {Sparveris}, \citenamefont {Stepanyan}, \citenamefont
  {Strakovsky}, \citenamefont {Strauch}, \citenamefont {Taiuti}, \citenamefont
  {Ungaro}, \citenamefont {Voskanyan}, \citenamefont {Voutier}, \citenamefont
  {Wei}, \citenamefont {Wood}, \citenamefont {Zachariou}, \citenamefont {Zana},
  \citenamefont {Zhang},\ and\ \citenamefont {Zhao}}]{Akbar2017Measurement}%
  \BibitemOpen
  \bibfield  {author} {\bibinfo {author} {\bibfnamefont {Z.}~\bibnamefont
  {Akbar}}, \bibinfo {author} {\bibfnamefont {P.}~\bibnamefont {Roy}}, \bibinfo
  {author} {\bibfnamefont {S.}~\bibnamefont {Park}}, \bibinfo {author}
  {\bibfnamefont {V.}~\bibnamefont {Crede}}, \bibinfo {author} {\bibfnamefont
  {A.~V.}\ \bibnamefont {Anisovich}}, \bibinfo {author} {\bibfnamefont
  {I.}~\bibnamefont {Denisenko}}, \bibinfo {author} {\bibfnamefont
  {E.}~\bibnamefont {Klempt}}, \bibinfo {author} {\bibfnamefont {V.~A.}\
  \bibnamefont {Nikonov}}, \bibinfo {author} {\bibfnamefont {A.~V.}\
  \bibnamefont {Sarantsev}}, \bibinfo {author} {\bibfnamefont {K.~P.}\
  \bibnamefont {Adhikari}}, \bibinfo {author} {\bibfnamefont {S.}~\bibnamefont
  {Adhikari}}, \bibinfo {author} {\bibfnamefont {M.~J.}\ \bibnamefont
  {Amaryan}}, \bibinfo {author} {\bibfnamefont {S.}~\bibnamefont
  {Anefalos~Pereira}}, \bibinfo {author} {\bibfnamefont {H.}~\bibnamefont
  {Avakian}}, \bibinfo {author} {\bibfnamefont {J.}~\bibnamefont {Ball}},
  \bibinfo {author} {\bibfnamefont {M.}~\bibnamefont {Battaglieri}}, \bibinfo
  {author} {\bibfnamefont {V.}~\bibnamefont {Batourine}}, \bibinfo {author}
  {\bibfnamefont {I.}~\bibnamefont {Bedlinskiy}}, \bibinfo {author}
  {\bibfnamefont {S.}~\bibnamefont {Boiarinov}}, \bibinfo {author}
  {\bibfnamefont {W.~J.}\ \bibnamefont {Briscoe}}, \bibinfo {author}
  {\bibfnamefont {J.}~\bibnamefont {Brock}}, \bibinfo {author} {\bibfnamefont
  {W.~K.}\ \bibnamefont {Brooks}}, \bibinfo {author} {\bibfnamefont {V.~D.}\
  \bibnamefont {Burkert}}, \bibinfo {author} {\bibfnamefont {F.~T.}\
  \bibnamefont {Cao}}, \bibinfo {author} {\bibfnamefont {C.}~\bibnamefont
  {Carlin}}, \bibinfo {author} {\bibfnamefont {D.~S.}\ \bibnamefont {Carman}},
  \bibinfo {author} {\bibfnamefont {A.}~\bibnamefont {Celentano}}, \bibinfo
  {author} {\bibfnamefont {G.}~\bibnamefont {Charles}}, \bibinfo {author}
  {\bibfnamefont {T.}~\bibnamefont {Chetry}}, \bibinfo {author} {\bibfnamefont
  {G.}~\bibnamefont {Ciullo}}, \bibinfo {author} {\bibfnamefont
  {L.}~\bibnamefont {Clark}}, \bibinfo {author} {\bibfnamefont {P.~L.}\
  \bibnamefont {Cole}}, \bibinfo {author} {\bibfnamefont {M.}~\bibnamefont
  {Contalbrigo}}, \bibinfo {author} {\bibfnamefont {O.}~\bibnamefont {Cortes}},
  \bibinfo {author} {\bibfnamefont {A.}~\bibnamefont {D'Angelo}}, \bibinfo
  {author} {\bibfnamefont {N.}~\bibnamefont {Dashyan}}, \bibinfo {author}
  {\bibfnamefont {R.}~\bibnamefont {De~Vita}}, \bibinfo {author} {\bibfnamefont
  {E.}~\bibnamefont {De~Sanctis}}, \bibinfo {author} {\bibfnamefont
  {A.}~\bibnamefont {Deur}}, \bibinfo {author} {\bibfnamefont {C.}~\bibnamefont
  {Djalali}}, \bibinfo {author} {\bibfnamefont {M.}~\bibnamefont {Dugger}},
  \bibinfo {author} {\bibfnamefont {R.}~\bibnamefont {Dupre}}, \bibinfo
  {author} {\bibfnamefont {H.}~\bibnamefont {Egiyan}}, \bibinfo {author}
  {\bibfnamefont {L.}~\bibnamefont {El~Fassi}}, \bibinfo {author}
  {\bibfnamefont {P.}~\bibnamefont {Eugenio}}, \bibinfo {author} {\bibfnamefont
  {G.}~\bibnamefont {Fedotov}}, \bibinfo {author} {\bibfnamefont
  {R.}~\bibnamefont {Fersch}}, \bibinfo {author} {\bibfnamefont
  {A.}~\bibnamefont {Filippi}}, \bibinfo {author} {\bibfnamefont
  {A.}~\bibnamefont {Fradi}}, \bibinfo {author} {\bibfnamefont
  {M.}~\bibnamefont {Gar\ifmmode~\mbox{\c{c}}\else \c{c}\fi{}on}}, \bibinfo
  {author} {\bibfnamefont {N.}~\bibnamefont {Gevorgyan}}, \bibinfo {author}
  {\bibfnamefont {K.~L.}\ \bibnamefont {Giovanetti}}, \bibinfo {author}
  {\bibfnamefont {F.~X.}\ \bibnamefont {Girod}}, \bibinfo {author}
  {\bibfnamefont {C.}~\bibnamefont {Gleason}}, \bibinfo {author} {\bibfnamefont
  {W.}~\bibnamefont {Gohn}}, \bibinfo {author} {\bibfnamefont {E.}~\bibnamefont
  {Golovatch}}, \bibinfo {author} {\bibfnamefont {R.~W.}\ \bibnamefont
  {Gothe}}, \bibinfo {author} {\bibfnamefont {K.~A.}\ \bibnamefont
  {Griffioen}}, \bibinfo {author} {\bibfnamefont {M.}~\bibnamefont {Guidal}},
  \bibinfo {author} {\bibfnamefont {L.}~\bibnamefont {Guo}}, \bibinfo {author}
  {\bibfnamefont {K.}~\bibnamefont {Hafidi}}, \bibinfo {author} {\bibfnamefont
  {H.}~\bibnamefont {Hakobyan}}, \bibinfo {author} {\bibfnamefont
  {C.}~\bibnamefont {Hanretty}}, \bibinfo {author} {\bibfnamefont
  {N.}~\bibnamefont {Harrison}}, \bibinfo {author} {\bibfnamefont
  {M.}~\bibnamefont {Hattawy}}, \bibinfo {author} {\bibfnamefont
  {D.}~\bibnamefont {Heddle}}, \bibinfo {author} {\bibfnamefont
  {K.}~\bibnamefont {Hicks}}, \bibinfo {author} {\bibfnamefont
  {G.}~\bibnamefont {Hollis}}, \bibinfo {author} {\bibfnamefont
  {M.}~\bibnamefont {Holtrop}}, \bibinfo {author} {\bibfnamefont {D.~G.}\
  \bibnamefont {Ireland}}, \bibinfo {author} {\bibfnamefont {B.~S.}\
  \bibnamefont {Ishkhanov}}, \bibinfo {author} {\bibfnamefont {E.~L.}\
  \bibnamefont {Isupov}}, \bibinfo {author} {\bibfnamefont {D.}~\bibnamefont
  {Jenkins}}, \bibinfo {author} {\bibfnamefont {S.}~\bibnamefont {Joosten}},
  \bibinfo {author} {\bibfnamefont {C.~D.}\ \bibnamefont {Keith}}, \bibinfo
  {author} {\bibfnamefont {D.}~\bibnamefont {Keller}}, \bibinfo {author}
  {\bibfnamefont {G.}~\bibnamefont {Khachatryan}}, \bibinfo {author}
  {\bibfnamefont {M.}~\bibnamefont {Khachatryan}}, \bibinfo {author}
  {\bibfnamefont {M.}~\bibnamefont {Khandaker}}, \bibinfo {author}
  {\bibfnamefont {A.}~\bibnamefont {Kim}}, \bibinfo {author} {\bibfnamefont
  {W.}~\bibnamefont {Kim}}, \bibinfo {author} {\bibfnamefont {A.}~\bibnamefont
  {Klein}}, \bibinfo {author} {\bibfnamefont {F.~J.}\ \bibnamefont {Klein}},
  \bibinfo {author} {\bibfnamefont {V.}~\bibnamefont {Kubarovsky}}, \bibinfo
  {author} {\bibfnamefont {L.}~\bibnamefont {Lanza}}, \bibinfo {author}
  {\bibfnamefont {K.}~\bibnamefont {Livingston}}, \bibinfo {author}
  {\bibfnamefont {I.~J.~D.}\ \bibnamefont {MacGregor}}, \bibinfo {author}
  {\bibfnamefont {N.}~\bibnamefont {Markov}}, \bibinfo {author} {\bibfnamefont
  {B.}~\bibnamefont {McKinnon}}, \bibinfo {author} {\bibfnamefont {D.~G.}\
  \bibnamefont {Meekins}}, \bibinfo {author} {\bibfnamefont {T.}~\bibnamefont
  {Mineeva}}, \bibinfo {author} {\bibfnamefont {V.}~\bibnamefont {Mokeev}},
  \bibinfo {author} {\bibfnamefont {A.}~\bibnamefont {Movsisyan}}, \bibinfo
  {author} {\bibfnamefont {C.}~\bibnamefont {Munoz~Camacho}}, \bibinfo {author}
  {\bibfnamefont {P.}~\bibnamefont {Nadel-Turonski}}, \bibinfo {author}
  {\bibfnamefont {S.}~\bibnamefont {Niccolai}}, \bibinfo {author}
  {\bibfnamefont {M.}~\bibnamefont {Osipenko}}, \bibinfo {author}
  {\bibfnamefont {A.~I.}\ \bibnamefont {Ostrovidov}}, \bibinfo {author}
  {\bibfnamefont {M.}~\bibnamefont {Paolone}}, \bibinfo {author} {\bibfnamefont
  {R.}~\bibnamefont {Paremuzyan}}, \bibinfo {author} {\bibfnamefont
  {K.}~\bibnamefont {Park}}, \bibinfo {author} {\bibfnamefont {E.}~\bibnamefont
  {Pasyuk}}, \bibinfo {author} {\bibfnamefont {W.}~\bibnamefont {Phelps}},
  \bibinfo {author} {\bibfnamefont {O.}~\bibnamefont {Pogorelko}}, \bibinfo
  {author} {\bibfnamefont {J.~W.}\ \bibnamefont {Price}}, \bibinfo {author}
  {\bibfnamefont {Y.}~\bibnamefont {Prok}}, \bibinfo {author} {\bibfnamefont
  {D.}~\bibnamefont {Protopopescu}}, \bibinfo {author} {\bibfnamefont {B.~A.}\
  \bibnamefont {Raue}}, \bibinfo {author} {\bibfnamefont {M.}~\bibnamefont
  {Ripani}}, \bibinfo {author} {\bibfnamefont {B.~G.}\ \bibnamefont {Ritchie}},
  \bibinfo {author} {\bibfnamefont {A.}~\bibnamefont {Rizzo}}, \bibinfo
  {author} {\bibfnamefont {G.}~\bibnamefont {Rosner}}, \bibinfo {author}
  {\bibfnamefont {F.}~\bibnamefont {Sabati\'e}}, \bibinfo {author}
  {\bibfnamefont {C.}~\bibnamefont {Salgado}}, \bibinfo {author} {\bibfnamefont
  {R.~A.}\ \bibnamefont {Schumacher}}, \bibinfo {author} {\bibfnamefont
  {Y.~G.}\ \bibnamefont {Sharabian}}, \bibinfo {author} {\bibfnamefont
  {I.}~\bibnamefont {Skorodumina}}, \bibinfo {author} {\bibfnamefont {G.~D.}\
  \bibnamefont {Smith}}, \bibinfo {author} {\bibfnamefont {D.~I.}\ \bibnamefont
  {Sober}}, \bibinfo {author} {\bibfnamefont {D.}~\bibnamefont {Sokhan}},
  \bibinfo {author} {\bibfnamefont {N.}~\bibnamefont {Sparveris}}, \bibinfo
  {author} {\bibfnamefont {S.}~\bibnamefont {Stepanyan}}, \bibinfo {author}
  {\bibfnamefont {I.~I.}\ \bibnamefont {Strakovsky}}, \bibinfo {author}
  {\bibfnamefont {S.}~\bibnamefont {Strauch}}, \bibinfo {author} {\bibfnamefont
  {M.}~\bibnamefont {Taiuti}}, \bibinfo {author} {\bibfnamefont
  {M.}~\bibnamefont {Ungaro}}, \bibinfo {author} {\bibfnamefont
  {H.}~\bibnamefont {Voskanyan}}, \bibinfo {author} {\bibfnamefont
  {E.}~\bibnamefont {Voutier}}, \bibinfo {author} {\bibfnamefont
  {X.}~\bibnamefont {Wei}}, \bibinfo {author} {\bibfnamefont {M.~H.}\
  \bibnamefont {Wood}}, \bibinfo {author} {\bibfnamefont {N.}~\bibnamefont
  {Zachariou}}, \bibinfo {author} {\bibfnamefont {L.}~\bibnamefont {Zana}},
  \bibinfo {author} {\bibfnamefont {J.}~\bibnamefont {Zhang}},\ and\ \bibinfo
  {author} {\bibfnamefont {Z.~W.}\ \bibnamefont {Zhao}} (\bibinfo
  {collaboration} {The CLAS Collaboration}),\ }\bibfield  {title} {\bibinfo
  {title} {{Measurement of the helicity asymmetry $E$ in
  $\ensuremath{\omega}\ensuremath{\rightarrow}{\ensuremath{\pi}}^{+}{\ensuremath{\pi}}^{\ensuremath{-}}{\ensuremath{\pi}}^{0}$
  photoproduction}},\ }\href {https://doi.org/10.1103/PhysRevC.96.065209}
  {\bibfield  {journal} {\bibinfo  {journal} {Phys. Rev. C}\ }\textbf {\bibinfo
  {volume} {96}},\ \bibinfo {pages} {065209} (\bibinfo {year}
  {2017})}\BibitemShut {NoStop}%
\bibitem [{\citenamefont {Krausz}\ and\ \citenamefont
  {Ivanov}(2009)}]{Krausz2009Atto}%
  \BibitemOpen
  \bibfield  {author} {\bibinfo {author} {\bibfnamefont {F.}~\bibnamefont
  {Krausz}}\ and\ \bibinfo {author} {\bibfnamefont {M.}~\bibnamefont
  {Ivanov}},\ }\bibfield  {title} {\bibinfo {title} {Attosecond physics},\
  }\href {https://doi.org/10.1103/RevModPhys.81.163} {\bibfield  {journal}
  {\bibinfo  {journal} {Rev. Mod. Phys.}\ }\textbf {\bibinfo {volume} {81}},\
  \bibinfo {pages} {163} (\bibinfo {year} {2009})}\BibitemShut {NoStop}%
\bibitem [{\citenamefont {Zilges}\ \emph {et~al.}(2022)\citenamefont {Zilges},
  \citenamefont {Balabanski}, \citenamefont {Isaak},\ and\ \citenamefont
  {Pietralla}}]{zilges2022photonuclear}%
  \BibitemOpen
  \bibfield  {author} {\bibinfo {author} {\bibfnamefont {A.}~\bibnamefont
  {Zilges}}, \bibinfo {author} {\bibfnamefont {D.}~\bibnamefont {Balabanski}},
  \bibinfo {author} {\bibfnamefont {J.}~\bibnamefont {Isaak}},\ and\ \bibinfo
  {author} {\bibfnamefont {N.}~\bibnamefont {Pietralla}},\ }\bibfield  {title}
  {\bibinfo {title} {Photonuclear reactions—from basic research to
  applications},\ }\href {https://doi.org/10.1016/j.ppnp.2021.103903}
  {\bibfield  {journal} {\bibinfo  {journal} {Prog. Part. Nucl. Phys.}\
  }\textbf {\bibinfo {volume} {122}},\ \bibinfo {pages} {103903} (\bibinfo
  {year} {2022})}\BibitemShut {NoStop}%
\bibitem [{\citenamefont {Li}\ \emph {et~al.}(2015)\citenamefont {Li},
  \citenamefont {Hatsagortsyan}, \citenamefont {Galow},\ and\ \citenamefont
  {Keitel}}]{li2015attosecond}%
  \BibitemOpen
  \bibfield  {author} {\bibinfo {author} {\bibfnamefont {J.-X.}\ \bibnamefont
  {Li}}, \bibinfo {author} {\bibfnamefont {K.~Z.}\ \bibnamefont
  {Hatsagortsyan}}, \bibinfo {author} {\bibfnamefont {B.~J.}\ \bibnamefont
  {Galow}},\ and\ \bibinfo {author} {\bibfnamefont {C.~H.}\ \bibnamefont
  {Keitel}},\ }\bibfield  {title} {\bibinfo {title} {{Attosecond Gamma-Ray
  Pulses via Nonlinear Compton Scattering in the Radiation-Dominated Regime}},\
  }\href {https://doi.org/10.1103/PhysRevLett.115.204801} {\bibfield  {journal}
  {\bibinfo  {journal} {Phys. Rev. Lett.}\ }\textbf {\bibinfo {volume} {115}},\
  \bibinfo {pages} {204801} (\bibinfo {year} {2015})}\BibitemShut {NoStop}%
\bibitem [{\citenamefont {Povh}\ \emph {et~al.}(1995)\citenamefont {Povh},
  \citenamefont {Rith}, \citenamefont {Scholz}, \citenamefont {Zetsche},\ and\
  \citenamefont {Rodejohann}}]{povh1995particles}%
  \BibitemOpen
  \bibfield  {author} {\bibinfo {author} {\bibfnamefont {B.}~\bibnamefont
  {Povh}}, \bibinfo {author} {\bibfnamefont {K.}~\bibnamefont {Rith}}, \bibinfo
  {author} {\bibfnamefont {C.}~\bibnamefont {Scholz}}, \bibinfo {author}
  {\bibfnamefont {F.}~\bibnamefont {Zetsche}},\ and\ \bibinfo {author}
  {\bibfnamefont {W.}~\bibnamefont {Rodejohann}},\ }\href
  {https://doi.org/10.1007/978-3-662-46321-5} {\emph {\bibinfo {title}
  {Particles and nuclei}}}\ (\bibinfo  {publisher} {Springer},\ \bibinfo {year}
  {1995})\BibitemShut {NoStop}%
\bibitem [{\citenamefont {Hentschel}\ \emph {et~al.}(2001)\citenamefont
  {Hentschel}, \citenamefont {Kienberger}, \citenamefont {Spielmann},
  \citenamefont {Reider}, \citenamefont {Milosevic}, \citenamefont {Brabec},
  \citenamefont {Corkum}, \citenamefont {Heinzmann}, \citenamefont {Drescher},\
  and\ \citenamefont {Krausz}}]{hentschel2001attosecond}%
  \BibitemOpen
  \bibfield  {author} {\bibinfo {author} {\bibfnamefont {M.}~\bibnamefont
  {Hentschel}}, \bibinfo {author} {\bibfnamefont {R.}~\bibnamefont
  {Kienberger}}, \bibinfo {author} {\bibfnamefont {C.}~\bibnamefont
  {Spielmann}}, \bibinfo {author} {\bibfnamefont {G.~A.}\ \bibnamefont
  {Reider}}, \bibinfo {author} {\bibfnamefont {N.}~\bibnamefont {Milosevic}},
  \bibinfo {author} {\bibfnamefont {T.}~\bibnamefont {Brabec}}, \bibinfo
  {author} {\bibfnamefont {P.}~\bibnamefont {Corkum}}, \bibinfo {author}
  {\bibfnamefont {U.}~\bibnamefont {Heinzmann}}, \bibinfo {author}
  {\bibfnamefont {M.}~\bibnamefont {Drescher}},\ and\ \bibinfo {author}
  {\bibfnamefont {F.}~\bibnamefont {Krausz}},\ }\bibfield  {title} {\bibinfo
  {title} {Attosecond metrology},\ }\href
  {https://doi.org/https://doi.org/10.1038/35107000} {\bibfield  {journal}
  {\bibinfo  {journal} {Nature}\ }\textbf {\bibinfo {volume} {414}},\ \bibinfo
  {pages} {509} (\bibinfo {year} {2001})}\BibitemShut {NoStop}%
\bibitem [{\citenamefont {Paul}\ \emph {et~al.}(2001)\citenamefont {Paul},
  \citenamefont {Toma}, \citenamefont {Breger}, \citenamefont {Mullot},
  \citenamefont {Augé}, \citenamefont {Balcou}, \citenamefont {Muller},\ and\
  \citenamefont {Agostini}}]{Paul2001Observation}%
  \BibitemOpen
  \bibfield  {author} {\bibinfo {author} {\bibfnamefont {P.~M.}\ \bibnamefont
  {Paul}}, \bibinfo {author} {\bibfnamefont {E.~S.}\ \bibnamefont {Toma}},
  \bibinfo {author} {\bibfnamefont {P.}~\bibnamefont {Breger}}, \bibinfo
  {author} {\bibfnamefont {G.}~\bibnamefont {Mullot}}, \bibinfo {author}
  {\bibfnamefont {F.}~\bibnamefont {Augé}}, \bibinfo {author} {\bibfnamefont
  {P.}~\bibnamefont {Balcou}}, \bibinfo {author} {\bibfnamefont {H.~G.}\
  \bibnamefont {Muller}},\ and\ \bibinfo {author} {\bibfnamefont
  {P.}~\bibnamefont {Agostini}},\ }\bibfield  {title} {\bibinfo {title}
  {{Observation of a Train of Attosecond Pulses from High Harmonic
  Generation}},\ }\href {https://doi.org/10.1126/science.1059413} {\bibfield
  {journal} {\bibinfo  {journal} {Science}\ }\textbf {\bibinfo {volume}
  {292}},\ \bibinfo {pages} {1689} (\bibinfo {year} {2001})}\BibitemShut
  {NoStop}%
\bibitem [{\citenamefont {Sansone}\ \emph {et~al.}(2011)\citenamefont
  {Sansone}, \citenamefont {Poletto},\ and\ \citenamefont
  {Nisoli}}]{sansone2011high}%
  \BibitemOpen
  \bibfield  {author} {\bibinfo {author} {\bibfnamefont {G.}~\bibnamefont
  {Sansone}}, \bibinfo {author} {\bibfnamefont {L.}~\bibnamefont {Poletto}},\
  and\ \bibinfo {author} {\bibfnamefont {M.}~\bibnamefont {Nisoli}},\
  }\bibfield  {title} {\bibinfo {title} {High-energy attosecond light
  sources},\ }\href {https://doi.org/https://doi.org/10.1038/nphoton.2011.167}
  {\bibfield  {journal} {\bibinfo  {journal} {Nat. Photonics}\ }\textbf
  {\bibinfo {volume} {5}},\ \bibinfo {pages} {655} (\bibinfo {year}
  {2011})}\BibitemShut {NoStop}%
\bibitem [{\citenamefont {Saldin}\ \emph {et~al.}(2004)\citenamefont {Saldin},
  \citenamefont {Schneidmiller},\ and\ \citenamefont {Yurkov}}]{Saldin2004TW}%
  \BibitemOpen
  \bibfield  {author} {\bibinfo {author} {\bibfnamefont {E.}~\bibnamefont
  {Saldin}}, \bibinfo {author} {\bibfnamefont {E.}~\bibnamefont
  {Schneidmiller}},\ and\ \bibinfo {author} {\bibfnamefont {M.}~\bibnamefont
  {Yurkov}},\ }\bibfield  {title} {\bibinfo {title} {{Terawatt-scale sub-10-fs
  laser technology – key to generation of {GW}-level attosecond pulses in
  {X}-ray free electron laser}},\ }\href
  {https://doi.org/https://doi.org/10.1016/j.optcom.2004.03.070} {\bibfield
  {journal} {\bibinfo  {journal} {Opt. Commun.}\ }\textbf {\bibinfo {volume}
  {237}},\ \bibinfo {pages} {153} (\bibinfo {year} {2004})}\BibitemShut
  {NoStop}%
\bibitem [{\citenamefont {Zholents}(2005)}]{Zholents2005Atto}%
  \BibitemOpen
  \bibfield  {author} {\bibinfo {author} {\bibfnamefont {A.~A.}\ \bibnamefont
  {Zholents}},\ }\bibfield  {title} {\bibinfo {title} {{Attosecond {X}-ray
  Pulses from Free-Electron Lasers}},\ }\href
  {https://www.researchgate.net/publication/255277556_Attosecond_X-ray_Pulses_from_Free-Electron_Lasers}
  {\bibfield  {journal} {\bibinfo  {journal} {Laser Phys.}\ }\textbf {\bibinfo
  {volume} {15}} (\bibinfo {year} {2005})}\BibitemShut {NoStop}%
\bibitem [{\citenamefont {Xu}\ \emph {et~al.}(2024)\citenamefont {Xu},
  \citenamefont {Liu}, \citenamefont {Dalichaouch}, \citenamefont {Tsung},
  \citenamefont {Zhang}, \citenamefont {Huang}, \citenamefont {Hogan},
  \citenamefont {Yan}, \citenamefont {Joshi},\ and\ \citenamefont
  {Mori}}]{xu2023ultracompact}%
  \BibitemOpen
  \bibfield  {author} {\bibinfo {author} {\bibfnamefont {X.}~\bibnamefont
  {Xu}}, \bibinfo {author} {\bibfnamefont {J.}~\bibnamefont {Liu}}, \bibinfo
  {author} {\bibfnamefont {T.}~\bibnamefont {Dalichaouch}}, \bibinfo {author}
  {\bibfnamefont {F.~S.}\ \bibnamefont {Tsung}}, \bibinfo {author}
  {\bibfnamefont {Z.}~\bibnamefont {Zhang}}, \bibinfo {author} {\bibfnamefont
  {Z.}~\bibnamefont {Huang}}, \bibinfo {author} {\bibfnamefont {M.~J.}\
  \bibnamefont {Hogan}}, \bibinfo {author} {\bibfnamefont {X.}~\bibnamefont
  {Yan}}, \bibinfo {author} {\bibfnamefont {C.}~\bibnamefont {Joshi}},\ and\
  \bibinfo {author} {\bibfnamefont {W.~B.}\ \bibnamefont {Mori}},\ }\bibfield
  {title} {\bibinfo {title} {{Attosecond x-ray free-electron lasers utilizing
  an optical undulator in a self-selection regime}},\ }\href
  {https://doi.org/10.1103/PhysRevAccelBeams.27.011301} {\bibfield  {journal}
  {\bibinfo  {journal} {Phys. Rev. Accel. Beams}\ }\textbf {\bibinfo {volume}
  {27}},\ \bibinfo {pages} {011301} (\bibinfo {year} {2024})}\BibitemShut
  {NoStop}%
\bibitem [{\citenamefont {Gu}\ \emph {et~al.}(2018)\citenamefont {Gu},
  \citenamefont {Klimo}, \citenamefont {Bulanov},\ and\ \citenamefont
  {Weber}}]{gu2018brilliant}%
  \BibitemOpen
  \bibfield  {author} {\bibinfo {author} {\bibfnamefont {Y.-J.}\ \bibnamefont
  {Gu}}, \bibinfo {author} {\bibfnamefont {O.}~\bibnamefont {Klimo}}, \bibinfo
  {author} {\bibfnamefont {S.~V.}\ \bibnamefont {Bulanov}},\ and\ \bibinfo
  {author} {\bibfnamefont {S.}~\bibnamefont {Weber}},\ }\bibfield  {title}
  {\bibinfo {title} {Brilliant gamma-ray beam and electron--positron pair
  production by enhanced attosecond pulses},\ }\href
  {https://doi.org/https://doi.org/10.1038/s42005-018-0095-3} {\bibfield
  {journal} {\bibinfo  {journal} {Commun. Phys.}\ }\textbf {\bibinfo {volume}
  {1}},\ \bibinfo {pages} {93} (\bibinfo {year} {2018})}\BibitemShut {NoStop}%
\bibitem [{\citenamefont {L\'ecz}\ and\ \citenamefont
  {Andreev}(2019)}]{lecz2019attosecond}%
  \BibitemOpen
  \bibfield  {author} {\bibinfo {author} {\bibfnamefont {Z.}~\bibnamefont
  {L\'ecz}}\ and\ \bibinfo {author} {\bibfnamefont {A.}~\bibnamefont
  {Andreev}},\ }\bibfield  {title} {\bibinfo {title} {Attosecond bunches of
  gamma photons and positrons generated in nanostructure targets},\ }\href
  {https://doi.org/10.1103/PhysRevE.99.013202} {\bibfield  {journal} {\bibinfo
  {journal} {Phys. Rev. E}\ }\textbf {\bibinfo {volume} {99}},\ \bibinfo
  {pages} {013202} (\bibinfo {year} {2019})}\BibitemShut {NoStop}%
\bibitem [{\citenamefont {Zhang}\ \emph {et~al.}(2021)\citenamefont {Zhang},
  \citenamefont {Wu}, \citenamefont {Huang}, \citenamefont {Lan}, \citenamefont
  {Liu}, \citenamefont {Wu}, \citenamefont {Yang}, \citenamefont {Zhao},
  \citenamefont {Zhu},\ and\ \citenamefont {Luo}}]{zhang2021brilliant}%
  \BibitemOpen
  \bibfield  {author} {\bibinfo {author} {\bibfnamefont {L.-Q.}\ \bibnamefont
  {Zhang}}, \bibinfo {author} {\bibfnamefont {S.-D.}\ \bibnamefont {Wu}},
  \bibinfo {author} {\bibfnamefont {H.-R.}\ \bibnamefont {Huang}}, \bibinfo
  {author} {\bibfnamefont {H.-Y.}\ \bibnamefont {Lan}}, \bibinfo {author}
  {\bibfnamefont {W.-Y.}\ \bibnamefont {Liu}}, \bibinfo {author} {\bibfnamefont
  {Y.-C.}\ \bibnamefont {Wu}}, \bibinfo {author} {\bibfnamefont
  {Y.}~\bibnamefont {Yang}}, \bibinfo {author} {\bibfnamefont {Z.-Q.}\
  \bibnamefont {Zhao}}, \bibinfo {author} {\bibfnamefont {Z.-C.}\ \bibnamefont
  {Zhu}},\ and\ \bibinfo {author} {\bibfnamefont {W.}~\bibnamefont {Luo}},\
  }\bibfield  {title} {\bibinfo {title} {{Brilliant attosecond $\gamma$-ray
  emission and high-yield positron production from intense laser-irradiated
  nano-micro array}},\ }\href {https://doi.org/10.1063/5.0030909} {\bibfield
  {journal} {\bibinfo  {journal} {Phys. Plasmas}\ }\textbf {\bibinfo {volume}
  {28}},\ \bibinfo {pages} {023110} (\bibinfo {year} {2021})}\BibitemShut
  {NoStop}%
\bibitem [{\citenamefont {Yanovsky}\ \emph {et~al.}(2008)\citenamefont
  {Yanovsky}, \citenamefont {Chvykov}, \citenamefont {Kalinchenko},
  \citenamefont {Rousseau}, \citenamefont {Planchon}, \citenamefont {Matsuoka},
  \citenamefont {Maksimchuk}, \citenamefont {Nees}, \citenamefont {Cheriaux},
  \citenamefont {Mourou},\ and\ \citenamefont
  {Krushelnick}}]{yanovsky2008ultra}%
  \BibitemOpen
  \bibfield  {author} {\bibinfo {author} {\bibfnamefont {V.}~\bibnamefont
  {Yanovsky}}, \bibinfo {author} {\bibfnamefont {V.}~\bibnamefont {Chvykov}},
  \bibinfo {author} {\bibfnamefont {G.}~\bibnamefont {Kalinchenko}}, \bibinfo
  {author} {\bibfnamefont {P.}~\bibnamefont {Rousseau}}, \bibinfo {author}
  {\bibfnamefont {T.}~\bibnamefont {Planchon}}, \bibinfo {author}
  {\bibfnamefont {T.}~\bibnamefont {Matsuoka}}, \bibinfo {author}
  {\bibfnamefont {A.}~\bibnamefont {Maksimchuk}}, \bibinfo {author}
  {\bibfnamefont {J.}~\bibnamefont {Nees}}, \bibinfo {author} {\bibfnamefont
  {G.}~\bibnamefont {Cheriaux}}, \bibinfo {author} {\bibfnamefont
  {G.}~\bibnamefont {Mourou}},\ and\ \bibinfo {author} {\bibfnamefont
  {K.}~\bibnamefont {Krushelnick}},\ }\bibfield  {title} {\bibinfo {title}
  {{Ultra-high intensity- 300-{TW} laser at 0.1 {Hz} repetition rate}},\ }\href
  {https://doi.org/10.1364/OE.16.002109} {\bibfield  {journal} {\bibinfo
  {journal} {Opt. Express}\ }\textbf {\bibinfo {volume} {16}},\ \bibinfo
  {pages} {2109} (\bibinfo {year} {2008})}\BibitemShut {NoStop}%
\bibitem [{\citenamefont {Pirozhkov}\ \emph {et~al.}(2017)\citenamefont
  {Pirozhkov}, \citenamefont {Fukuda}, \citenamefont {Nishiuchi}, \citenamefont
  {Kiriyama}, \citenamefont {Sagisaka}, \citenamefont {Ogura}, \citenamefont
  {Mori}, \citenamefont {Kishimoto}, \citenamefont {Sakaki}, \citenamefont
  {Dover}, \citenamefont {Kondo}, \citenamefont {Nakanii}, \citenamefont
  {Huang}, \citenamefont {Kanasaki}, \citenamefont {Kondo},\ and\ \citenamefont
  {Kando}}]{pirozhkov2017approaching}%
  \BibitemOpen
  \bibfield  {author} {\bibinfo {author} {\bibfnamefont {A.~S.}\ \bibnamefont
  {Pirozhkov}}, \bibinfo {author} {\bibfnamefont {Y.}~\bibnamefont {Fukuda}},
  \bibinfo {author} {\bibfnamefont {M.}~\bibnamefont {Nishiuchi}}, \bibinfo
  {author} {\bibfnamefont {H.}~\bibnamefont {Kiriyama}}, \bibinfo {author}
  {\bibfnamefont {A.}~\bibnamefont {Sagisaka}}, \bibinfo {author}
  {\bibfnamefont {K.}~\bibnamefont {Ogura}}, \bibinfo {author} {\bibfnamefont
  {M.}~\bibnamefont {Mori}}, \bibinfo {author} {\bibfnamefont {M.}~\bibnamefont
  {Kishimoto}}, \bibinfo {author} {\bibfnamefont {H.}~\bibnamefont {Sakaki}},
  \bibinfo {author} {\bibfnamefont {N.~P.}\ \bibnamefont {Dover}}, \bibinfo
  {author} {\bibfnamefont {K.}~\bibnamefont {Kondo}}, \bibinfo {author}
  {\bibfnamefont {N.}~\bibnamefont {Nakanii}}, \bibinfo {author} {\bibfnamefont
  {K.}~\bibnamefont {Huang}}, \bibinfo {author} {\bibfnamefont
  {M.}~\bibnamefont {Kanasaki}}, \bibinfo {author} {\bibfnamefont
  {K.}~\bibnamefont {Kondo}},\ and\ \bibinfo {author} {\bibfnamefont
  {M.}~\bibnamefont {Kando}},\ }\bibfield  {title} {\bibinfo {title}
  {{Approaching the diffraction-limited, bandwidth-limited {P}etawatt}},\
  }\href {https://doi.org/10.1364/OE.25.020486} {\bibfield  {journal} {\bibinfo
   {journal} {Opt. Express}\ }\textbf {\bibinfo {volume} {25}},\ \bibinfo
  {pages} {20486} (\bibinfo {year} {2017})}\BibitemShut {NoStop}%
\bibitem [{\citenamefont {Guo}\ \emph {et~al.}(2018)\citenamefont {Guo},
  \citenamefont {Yu}, \citenamefont {Wang}, \citenamefont {Wang}, \citenamefont
  {Liu}, \citenamefont {Gan}, \citenamefont {Li}, \citenamefont {Leng},
  \citenamefont {Liang},\ and\ \citenamefont {Li}}]{guo2018improvement}%
  \BibitemOpen
  \bibfield  {author} {\bibinfo {author} {\bibfnamefont {Z.}~\bibnamefont
  {Guo}}, \bibinfo {author} {\bibfnamefont {L.}~\bibnamefont {Yu}}, \bibinfo
  {author} {\bibfnamefont {J.}~\bibnamefont {Wang}}, \bibinfo {author}
  {\bibfnamefont {C.}~\bibnamefont {Wang}}, \bibinfo {author} {\bibfnamefont
  {Y.}~\bibnamefont {Liu}}, \bibinfo {author} {\bibfnamefont {Z.}~\bibnamefont
  {Gan}}, \bibinfo {author} {\bibfnamefont {W.}~\bibnamefont {Li}}, \bibinfo
  {author} {\bibfnamefont {Y.}~\bibnamefont {Leng}}, \bibinfo {author}
  {\bibfnamefont {X.}~\bibnamefont {Liang}},\ and\ \bibinfo {author}
  {\bibfnamefont {R.}~\bibnamefont {Li}},\ }\bibfield  {title} {\bibinfo
  {title} {{Improvement of the focusing ability by double deformable mirrors
  for 10-{PW}-level {T}i: sapphire chirped pulse amplification laser system}},\
  }\href {https://doi.org/10.1364/OE.26.026776} {\bibfield  {journal} {\bibinfo
   {journal} {Opt. Express}\ }\textbf {\bibinfo {volume} {26}},\ \bibinfo
  {pages} {26776} (\bibinfo {year} {2018})}\BibitemShut {NoStop}%
\bibitem [{\citenamefont {Tiwari}\ \emph {et~al.}(2019)\citenamefont {Tiwari},
  \citenamefont {Gaul}, \citenamefont {Martinez}, \citenamefont {Dyer},
  \citenamefont {Gordon}, \citenamefont {Spinks}, \citenamefont {Toncian},
  \citenamefont {Bowers}, \citenamefont {Jiao}, \citenamefont {Kupfer},
  \citenamefont {Lisi}, \citenamefont {McCary}, \citenamefont {Roycroft},
  \citenamefont {Yandow}, \citenamefont {Glenn}, \citenamefont {Donovan},
  \citenamefont {Ditmire},\ and\ \citenamefont {Hegelich}}]{tiwari2019beam}%
  \BibitemOpen
  \bibfield  {author} {\bibinfo {author} {\bibfnamefont {G.}~\bibnamefont
  {Tiwari}}, \bibinfo {author} {\bibfnamefont {E.}~\bibnamefont {Gaul}},
  \bibinfo {author} {\bibfnamefont {M.}~\bibnamefont {Martinez}}, \bibinfo
  {author} {\bibfnamefont {G.}~\bibnamefont {Dyer}}, \bibinfo {author}
  {\bibfnamefont {J.}~\bibnamefont {Gordon}}, \bibinfo {author} {\bibfnamefont
  {M.}~\bibnamefont {Spinks}}, \bibinfo {author} {\bibfnamefont
  {T.}~\bibnamefont {Toncian}}, \bibinfo {author} {\bibfnamefont
  {B.}~\bibnamefont {Bowers}}, \bibinfo {author} {\bibfnamefont
  {X.}~\bibnamefont {Jiao}}, \bibinfo {author} {\bibfnamefont {R.}~\bibnamefont
  {Kupfer}}, \bibinfo {author} {\bibfnamefont {L.}~\bibnamefont {Lisi}},
  \bibinfo {author} {\bibfnamefont {E.}~\bibnamefont {McCary}}, \bibinfo
  {author} {\bibfnamefont {R.}~\bibnamefont {Roycroft}}, \bibinfo {author}
  {\bibfnamefont {A.}~\bibnamefont {Yandow}}, \bibinfo {author} {\bibfnamefont
  {G.~D.}\ \bibnamefont {Glenn}}, \bibinfo {author} {\bibfnamefont
  {M.}~\bibnamefont {Donovan}}, \bibinfo {author} {\bibfnamefont
  {T.}~\bibnamefont {Ditmire}},\ and\ \bibinfo {author} {\bibfnamefont {B.~M.}\
  \bibnamefont {Hegelich}},\ }\bibfield  {title} {\bibinfo {title} {{Beam
  distortion effects upon focusing an ultrashort petawatt laser pulse to
  greater than 10$^{22}${W}/cm$^{2}$}},\ }\href
  {https://doi.org/10.1364/OL.44.002764} {\bibfield  {journal} {\bibinfo
  {journal} {Opt. Lett.}\ }\textbf {\bibinfo {volume} {44}},\ \bibinfo {pages}
  {2764} (\bibinfo {year} {2019})}\BibitemShut {NoStop}%
\bibitem [{\citenamefont {Yoon}\ \emph {et~al.}(2019)\citenamefont {Yoon},
  \citenamefont {Jeon}, \citenamefont {Shin}, \citenamefont {Lee},
  \citenamefont {Lee}, \citenamefont {Choi}, \citenamefont {Kim}, \citenamefont
  {Sung},\ and\ \citenamefont {Nam}}]{yoon2019achieving}%
  \BibitemOpen
  \bibfield  {author} {\bibinfo {author} {\bibfnamefont {J.~W.}\ \bibnamefont
  {Yoon}}, \bibinfo {author} {\bibfnamefont {C.}~\bibnamefont {Jeon}}, \bibinfo
  {author} {\bibfnamefont {J.}~\bibnamefont {Shin}}, \bibinfo {author}
  {\bibfnamefont {S.~K.}\ \bibnamefont {Lee}}, \bibinfo {author} {\bibfnamefont
  {H.~W.}\ \bibnamefont {Lee}}, \bibinfo {author} {\bibfnamefont {I.~W.}\
  \bibnamefont {Choi}}, \bibinfo {author} {\bibfnamefont {H.~T.}\ \bibnamefont
  {Kim}}, \bibinfo {author} {\bibfnamefont {J.~H.}\ \bibnamefont {Sung}},\ and\
  \bibinfo {author} {\bibfnamefont {C.~H.}\ \bibnamefont {Nam}},\ }\bibfield
  {title} {\bibinfo {title} {{Achieving the laser intensity of 5.5$\times$
  10$^{22}${W}/cm$^{2}$ with a wavefront-corrected multi-{PW} laser}},\ }\href
  {https://doi.org/10.1364/OE.27.020412} {\bibfield  {journal} {\bibinfo
  {journal} {Opt. Express}\ }\textbf {\bibinfo {volume} {27}},\ \bibinfo
  {pages} {20412} (\bibinfo {year} {2019})}\BibitemShut {NoStop}%
\bibitem [{\citenamefont {Yoon}\ \emph {et~al.}(2021)\citenamefont {Yoon},
  \citenamefont {Kim}, \citenamefont {Choi}, \citenamefont {Sung},
  \citenamefont {Lee}, \citenamefont {Lee},\ and\ \citenamefont
  {Nam}}]{yoon2021realization}%
  \BibitemOpen
  \bibfield  {author} {\bibinfo {author} {\bibfnamefont {J.~W.}\ \bibnamefont
  {Yoon}}, \bibinfo {author} {\bibfnamefont {Y.~G.}\ \bibnamefont {Kim}},
  \bibinfo {author} {\bibfnamefont {I.~W.}\ \bibnamefont {Choi}}, \bibinfo
  {author} {\bibfnamefont {J.~H.}\ \bibnamefont {Sung}}, \bibinfo {author}
  {\bibfnamefont {H.~W.}\ \bibnamefont {Lee}}, \bibinfo {author} {\bibfnamefont
  {S.~K.}\ \bibnamefont {Lee}},\ and\ \bibinfo {author} {\bibfnamefont {C.~H.}\
  \bibnamefont {Nam}},\ }\bibfield  {title} {\bibinfo {title} {{Realization of
  laser intensity over 10$^{23}${W}/cm$^{2}$}},\ }\href
  {https://doi.org/10.1364/OPTICA.420520} {\bibfield  {journal} {\bibinfo
  {journal} {Optica}\ }\textbf {\bibinfo {volume} {8}},\ \bibinfo {pages} {630}
  (\bibinfo {year} {2021})}\BibitemShut {NoStop}%
\bibitem [{\citenamefont {Li}\ \emph {et~al.}(2017)\citenamefont {Li},
  \citenamefont {Yu}, \citenamefont {Hu}, \citenamefont {Yin}, \citenamefont
  {Zou}, \citenamefont {Liu}, \citenamefont {Wang}, \citenamefont {Hu},\ and\
  \citenamefont {Shao}}]{li2017ultra}%
  \BibitemOpen
  \bibfield  {author} {\bibinfo {author} {\bibfnamefont {H.-Z.}\ \bibnamefont
  {Li}}, \bibinfo {author} {\bibfnamefont {T.-P.}\ \bibnamefont {Yu}}, \bibinfo
  {author} {\bibfnamefont {L.-X.}\ \bibnamefont {Hu}}, \bibinfo {author}
  {\bibfnamefont {Y.}~\bibnamefont {Yin}}, \bibinfo {author} {\bibfnamefont
  {D.-B.}\ \bibnamefont {Zou}}, \bibinfo {author} {\bibfnamefont {J.-X.}\
  \bibnamefont {Liu}}, \bibinfo {author} {\bibfnamefont {W.-Q.}\ \bibnamefont
  {Wang}}, \bibinfo {author} {\bibfnamefont {S.}~\bibnamefont {Hu}},\ and\
  \bibinfo {author} {\bibfnamefont {F.-Q.}\ \bibnamefont {Shao}},\ }\bibfield
  {title} {\bibinfo {title} {{Ultra-bright $\gamma$-ray flashes and dense
  attosecond positron bunches from two counter-propagating laser pulses
  irradiating a micro-wire target}},\ }\href
  {https://doi.org/10.1364/OE.25.021583} {\bibfield  {journal} {\bibinfo
  {journal} {Opt. Express}\ }\textbf {\bibinfo {volume} {25}},\ \bibinfo
  {pages} {21583} (\bibinfo {year} {2017})}\BibitemShut {NoStop}%
\bibitem [{\citenamefont {Zhang}\ \emph {et~al.}(2022)\citenamefont {Zhang},
  \citenamefont {Liu}, \citenamefont {Tang}, \citenamefont {Luo}, \citenamefont
  {Zhao}, \citenamefont {Zhang},\ and\ \citenamefont
  {Yu}}]{zhang2022generation}%
  \BibitemOpen
  \bibfield  {author} {\bibinfo {author} {\bibfnamefont {L.-Q.}\ \bibnamefont
  {Zhang}}, \bibinfo {author} {\bibfnamefont {K.}~\bibnamefont {Liu}}, \bibinfo
  {author} {\bibfnamefont {S.}~\bibnamefont {Tang}}, \bibinfo {author}
  {\bibfnamefont {W.}~\bibnamefont {Luo}}, \bibinfo {author} {\bibfnamefont
  {J.}~\bibnamefont {Zhao}}, \bibinfo {author} {\bibfnamefont {H.}~\bibnamefont
  {Zhang}},\ and\ \bibinfo {author} {\bibfnamefont {T.-P.}\ \bibnamefont
  {Yu}},\ }\bibfield  {title} {\bibinfo {title} {Generation of isolated and
  polarized $\gamma$-ray pulse by few-cycle laser irradiating a nanofoil},\
  }\href {https://doi.org/10.1088/1361-6587/ac85a7} {\bibfield  {journal}
  {\bibinfo  {journal} {Plasma Phys. Control. Fusion}\ }\textbf {\bibinfo
  {volume} {64}},\ \bibinfo {pages} {105011} (\bibinfo {year}
  {2022})}\BibitemShut {NoStop}%
\bibitem [{\citenamefont {Zhu}\ \emph {et~al.}(2018)\citenamefont {Zhu},
  \citenamefont {Chen}, \citenamefont {Yu}, \citenamefont {Weng}, \citenamefont
  {Hu}, \citenamefont {McKenna},\ and\ \citenamefont {Sheng}}]{zhu2018bright}%
  \BibitemOpen
  \bibfield  {author} {\bibinfo {author} {\bibfnamefont {X.-L.}\ \bibnamefont
  {Zhu}}, \bibinfo {author} {\bibfnamefont {M.}~\bibnamefont {Chen}}, \bibinfo
  {author} {\bibfnamefont {T.-P.}\ \bibnamefont {Yu}}, \bibinfo {author}
  {\bibfnamefont {S.-M.}\ \bibnamefont {Weng}}, \bibinfo {author}
  {\bibfnamefont {L.-X.}\ \bibnamefont {Hu}}, \bibinfo {author} {\bibfnamefont
  {P.}~\bibnamefont {McKenna}},\ and\ \bibinfo {author} {\bibfnamefont {Z.-M.}\
  \bibnamefont {Sheng}},\ }\bibfield  {title} {\bibinfo {title} {{Bright
  attosecond $\gamma$-ray pulses from nonlinear Compton scattering with
  laser-illuminated compound targets}},\ }\href
  {https://doi.org/10.1063/1.5028555} {\bibfield  {journal} {\bibinfo
  {journal} {Appl. Phys. Lett.}\ }\textbf {\bibinfo {volume} {112}},\ \bibinfo
  {pages} {174102} (\bibinfo {year} {2018})}\BibitemShut {NoStop}%
\bibitem [{\citenamefont {Zhu}\ \emph {et~al.}(2019)\citenamefont {Zhu},
  \citenamefont {Chen}, \citenamefont {Yu}, \citenamefont {Weng}, \citenamefont
  {He},\ and\ \citenamefont {Sheng}}]{zhu2019collimated}%
  \BibitemOpen
  \bibfield  {author} {\bibinfo {author} {\bibfnamefont {X.-L.}\ \bibnamefont
  {Zhu}}, \bibinfo {author} {\bibfnamefont {M.}~\bibnamefont {Chen}}, \bibinfo
  {author} {\bibfnamefont {T.-P.}\ \bibnamefont {Yu}}, \bibinfo {author}
  {\bibfnamefont {S.-M.}\ \bibnamefont {Weng}}, \bibinfo {author}
  {\bibfnamefont {F.}~\bibnamefont {He}},\ and\ \bibinfo {author}
  {\bibfnamefont {Z.-M.}\ \bibnamefont {Sheng}},\ }\bibfield  {title} {\bibinfo
  {title} {{{Collimated GeV attosecond electron–positron bunches from a
  plasma channel driven by 10 PW lasers}}},\ }\href
  {https://doi.org/10.1063/1.5083914} {\bibfield  {journal} {\bibinfo
  {journal} {Matter Radiat. Extremes}\ }\textbf {\bibinfo {volume} {4}},\
  \bibinfo {pages} {014401} (\bibinfo {year} {2019})}\BibitemShut {NoStop}%
\bibitem [{\citenamefont {Hu}\ \emph {et~al.}(2021)\citenamefont {Hu},
  \citenamefont {Zhao}, \citenamefont {Zhang}, \citenamefont {Lu},
  \citenamefont {Wang}, \citenamefont {Hu}, \citenamefont {Shao},\ and\
  \citenamefont {Yu}}]{Hu2021Atto}%
  \BibitemOpen
  \bibfield  {author} {\bibinfo {author} {\bibfnamefont {Y.-T.}\ \bibnamefont
  {Hu}}, \bibinfo {author} {\bibfnamefont {J.}~\bibnamefont {Zhao}}, \bibinfo
  {author} {\bibfnamefont {H.}~\bibnamefont {Zhang}}, \bibinfo {author}
  {\bibfnamefont {Y.}~\bibnamefont {Lu}}, \bibinfo {author} {\bibfnamefont
  {W.-Q.}\ \bibnamefont {Wang}}, \bibinfo {author} {\bibfnamefont {L.-X.}\
  \bibnamefont {Hu}}, \bibinfo {author} {\bibfnamefont {F.-Q.}\ \bibnamefont
  {Shao}},\ and\ \bibinfo {author} {\bibfnamefont {T.-P.}\ \bibnamefont {Yu}},\
  }\bibfield  {title} {\bibinfo {title} {{Attosecond $\gamma$-ray vortex
  generation in near-critical-density plasma driven by twisted laser pulses}},\
  }\href {https://doi.org/10.1063/5.0028203} {\bibfield  {journal} {\bibinfo
  {journal} {Appl. Phys. Lett.}\ }\textbf {\bibinfo {volume} {118}},\ \bibinfo
  {pages} {054101} (\bibinfo {year} {2021})}\BibitemShut {NoStop}%
\bibitem [{\citenamefont {Kumar}\ \emph {et~al.}(2010)\citenamefont {Kumar},
  \citenamefont {Pukhov},\ and\ \citenamefont
  {Lotov}}]{Kumar2010Self-modulation}%
  \BibitemOpen
  \bibfield  {author} {\bibinfo {author} {\bibfnamefont {N.}~\bibnamefont
  {Kumar}}, \bibinfo {author} {\bibfnamefont {A.}~\bibnamefont {Pukhov}},\ and\
  \bibinfo {author} {\bibfnamefont {K.}~\bibnamefont {Lotov}},\ }\bibfield
  {title} {\bibinfo {title} {{Self-Modulation Instability of a Long Proton
  Bunch in Plasmas}},\ }\href {https://doi.org/10.1103/PhysRevLett.104.255003}
  {\bibfield  {journal} {\bibinfo  {journal} {Phys. Rev. Lett.}\ }\textbf
  {\bibinfo {volume} {104}},\ \bibinfo {pages} {255003} (\bibinfo {year}
  {2010})}\BibitemShut {NoStop}%
\bibitem [{\citenamefont {Vieira}\ \emph {et~al.}(2012)\citenamefont {Vieira},
  \citenamefont {Fang}, \citenamefont {Mori}, \citenamefont {Silva},\ and\
  \citenamefont {Muggli}}]{Vieira2012Transverse}%
  \BibitemOpen
  \bibfield  {author} {\bibinfo {author} {\bibfnamefont {J.}~\bibnamefont
  {Vieira}}, \bibinfo {author} {\bibfnamefont {Y.}~\bibnamefont {Fang}},
  \bibinfo {author} {\bibfnamefont {W.~B.}\ \bibnamefont {Mori}}, \bibinfo
  {author} {\bibfnamefont {L.~O.}\ \bibnamefont {Silva}},\ and\ \bibinfo
  {author} {\bibfnamefont {P.}~\bibnamefont {Muggli}},\ }\bibfield  {title}
  {\bibinfo {title} {{Transverse self-modulation of ultra-relativistic lepton
  beams in the plasma wakefield accelerator}},\ }\href
  {https://doi.org/10.1063/1.4725425} {\bibfield  {journal} {\bibinfo
  {journal} {Phys. Plasmas}\ }\textbf {\bibinfo {volume} {19}},\ \bibinfo
  {pages} {063105} (\bibinfo {year} {2012})}\BibitemShut {NoStop}%
\bibitem [{\citenamefont {Benedetti}\ \emph {et~al.}(2018)\citenamefont
  {Benedetti}, \citenamefont {Tamburini},\ and\ \citenamefont
  {Keitel}}]{benedetti2018giant}%
  \BibitemOpen
  \bibfield  {author} {\bibinfo {author} {\bibfnamefont {A.}~\bibnamefont
  {Benedetti}}, \bibinfo {author} {\bibfnamefont {M.}~\bibnamefont
  {Tamburini}},\ and\ \bibinfo {author} {\bibfnamefont {C.~H.}\ \bibnamefont
  {Keitel}},\ }\bibfield  {title} {\bibinfo {title} {{Giant collimated
  gamma-ray flashes}},\ }\href
  {https://doi.org/https://doi.org/10.1038/s41566-018-0139-y} {\bibfield
  {journal} {\bibinfo  {journal} {Nat. Photonics}\ }\textbf {\bibinfo {volume}
  {12}},\ \bibinfo {pages} {319} (\bibinfo {year} {2018})}\BibitemShut
  {NoStop}%
\bibitem [{\citenamefont {Sampath}\ \emph {et~al.}(2021)\citenamefont
  {Sampath}, \citenamefont {Davoine}, \citenamefont {Corde}, \citenamefont
  {Gremillet}, \citenamefont {Gilljohann}, \citenamefont {Sangal},
  \citenamefont {Keitel}, \citenamefont {Ariniello}, \citenamefont {Cary},
  \citenamefont {Ekerfelt}, \citenamefont {Emma}, \citenamefont {Fiuza},
  \citenamefont {Fujii}, \citenamefont {Hogan}, \citenamefont {Joshi},
  \citenamefont {Knetsch}, \citenamefont {Kononenko}, \citenamefont {Lee},
  \citenamefont {Litos}, \citenamefont {Marsh}, \citenamefont {Nie},
  \citenamefont {O'Shea}, \citenamefont {Peterson}, \citenamefont {Claveria},
  \citenamefont {Storey}, \citenamefont {Wu}, \citenamefont {Xu}, \citenamefont
  {Zhang},\ and\ \citenamefont {Tamburini}}]{Sampath2021Extremely}%
  \BibitemOpen
  \bibfield  {author} {\bibinfo {author} {\bibfnamefont {A.}~\bibnamefont
  {Sampath}}, \bibinfo {author} {\bibfnamefont {X.}~\bibnamefont {Davoine}},
  \bibinfo {author} {\bibfnamefont {S.}~\bibnamefont {Corde}}, \bibinfo
  {author} {\bibfnamefont {L.}~\bibnamefont {Gremillet}}, \bibinfo {author}
  {\bibfnamefont {M.}~\bibnamefont {Gilljohann}}, \bibinfo {author}
  {\bibfnamefont {M.}~\bibnamefont {Sangal}}, \bibinfo {author} {\bibfnamefont
  {C.~H.}\ \bibnamefont {Keitel}}, \bibinfo {author} {\bibfnamefont
  {R.}~\bibnamefont {Ariniello}}, \bibinfo {author} {\bibfnamefont
  {J.}~\bibnamefont {Cary}}, \bibinfo {author} {\bibfnamefont {H.}~\bibnamefont
  {Ekerfelt}}, \bibinfo {author} {\bibfnamefont {C.}~\bibnamefont {Emma}},
  \bibinfo {author} {\bibfnamefont {F.}~\bibnamefont {Fiuza}}, \bibinfo
  {author} {\bibfnamefont {H.}~\bibnamefont {Fujii}}, \bibinfo {author}
  {\bibfnamefont {M.}~\bibnamefont {Hogan}}, \bibinfo {author} {\bibfnamefont
  {C.}~\bibnamefont {Joshi}}, \bibinfo {author} {\bibfnamefont
  {A.}~\bibnamefont {Knetsch}}, \bibinfo {author} {\bibfnamefont
  {O.}~\bibnamefont {Kononenko}}, \bibinfo {author} {\bibfnamefont
  {V.}~\bibnamefont {Lee}}, \bibinfo {author} {\bibfnamefont {M.}~\bibnamefont
  {Litos}}, \bibinfo {author} {\bibfnamefont {K.}~\bibnamefont {Marsh}},
  \bibinfo {author} {\bibfnamefont {Z.}~\bibnamefont {Nie}}, \bibinfo {author}
  {\bibfnamefont {B.}~\bibnamefont {O'Shea}}, \bibinfo {author} {\bibfnamefont
  {J.~R.}\ \bibnamefont {Peterson}}, \bibinfo {author} {\bibfnamefont
  {P.~S.~M.}\ \bibnamefont {Claveria}}, \bibinfo {author} {\bibfnamefont
  {D.}~\bibnamefont {Storey}}, \bibinfo {author} {\bibfnamefont
  {Y.}~\bibnamefont {Wu}}, \bibinfo {author} {\bibfnamefont {X.}~\bibnamefont
  {Xu}}, \bibinfo {author} {\bibfnamefont {C.}~\bibnamefont {Zhang}},\ and\
  \bibinfo {author} {\bibfnamefont {M.}~\bibnamefont {Tamburini}},\ }\bibfield
  {title} {\bibinfo {title} {{Extremely Dense Gamma-Ray Pulses in Electron
  Beam-Multifoil Collisions}},\ }\href
  {https://doi.org/10.1103/PhysRevLett.126.064801} {\bibfield  {journal}
  {\bibinfo  {journal} {Phys. Rev. Lett.}\ }\textbf {\bibinfo {volume} {126}},\
  \bibinfo {pages} {064801} (\bibinfo {year} {2021})}\BibitemShut {NoStop}%
\bibitem [{\citenamefont {Wang}\ \emph {et~al.}(2019)\citenamefont {Wang},
  \citenamefont {Feng}, \citenamefont {Liu}, \citenamefont {Zhang},
  \citenamefont {Tsai}, \citenamefont {Wu}, \citenamefont {Yang},\ and\
  \citenamefont {Zhao}}]{wang2019angular}%
  \BibitemOpen
  \bibfield  {author} {\bibinfo {author} {\bibfnamefont {X.}~\bibnamefont
  {Wang}}, \bibinfo {author} {\bibfnamefont {C.}~\bibnamefont {Feng}}, \bibinfo
  {author} {\bibfnamefont {T.}~\bibnamefont {Liu}}, \bibinfo {author}
  {\bibfnamefont {Z.}~\bibnamefont {Zhang}}, \bibinfo {author} {\bibfnamefont
  {C.-Y.}\ \bibnamefont {Tsai}}, \bibinfo {author} {\bibfnamefont
  {J.}~\bibnamefont {Wu}}, \bibinfo {author} {\bibfnamefont {C.}~\bibnamefont
  {Yang}},\ and\ \bibinfo {author} {\bibfnamefont {Z.}~\bibnamefont {Zhao}},\
  }\bibfield  {title} {\bibinfo {title} {{Angular dispersion enhanced prebunch
  for seeding ultrashort and coherent EUV and soft X-ray free-electron laser in
  storage rings}},\ }\href {https://doi.org/10.1107/S1600577519002674}
  {\bibfield  {journal} {\bibinfo  {journal} {J. Synchrotron Rad.}\ }\textbf
  {\bibinfo {volume} {26}},\ \bibinfo {pages} {677} (\bibinfo {year}
  {2019})}\BibitemShut {NoStop}%
\bibitem [{Sup()}]{Supplemental-Materials}%
  \BibitemOpen
  \href@noop {} {\bibinfo  {journal} {Supplemental Materials mainly include the
  high-quality electron microbunches, the modulation of ultra-relativistic
  electron beam and the description of photon polarization}\ }\BibitemShut
  {NoStop}%
\bibitem [{\citenamefont {Wan}\ \emph {et~al.}(2023)\citenamefont {Wan},
  \citenamefont {Lv}, \citenamefont {Xue}, \citenamefont {Dou}, \citenamefont
  {Zhao}, \citenamefont {Ababekri}, \citenamefont {Wei}, \citenamefont {Li},
  \citenamefont {Zhao},\ and\ \citenamefont {Li}}]{wan2023Simulations}%
  \BibitemOpen
\bibfield  {journal} {  }\bibfield  {author} {\bibinfo {author} {\bibfnamefont
  {F.}~\bibnamefont {Wan}}, \bibinfo {author} {\bibfnamefont {C.}~\bibnamefont
  {Lv}}, \bibinfo {author} {\bibfnamefont {K.}~\bibnamefont {Xue}}, \bibinfo
  {author} {\bibfnamefont {Z.-K.}\ \bibnamefont {Dou}}, \bibinfo {author}
  {\bibfnamefont {Q.}~\bibnamefont {Zhao}}, \bibinfo {author} {\bibfnamefont
  {M.}~\bibnamefont {Ababekri}}, \bibinfo {author} {\bibfnamefont {W.-Q.}\
  \bibnamefont {Wei}}, \bibinfo {author} {\bibfnamefont {Z.-P.}\ \bibnamefont
  {Li}}, \bibinfo {author} {\bibfnamefont {Y.-T.}\ \bibnamefont {Zhao}},\ and\
  \bibinfo {author} {\bibfnamefont {J.-X.}\ \bibnamefont {Li}},\ }\bibfield
  {title} {\bibinfo {title} {{Simulations of spin/polarization-resolved
  laser–plasma interactions in the nonlinear QED regime}},\ }\href
  {https://doi.org/10.1063/5.0163929} {\bibfield  {journal} {\bibinfo
  {journal} {Matter Radiat. Extremes}\ }\textbf {\bibinfo {volume} {8}},\
  \bibinfo {pages} {064002} (\bibinfo {year} {2023})}\BibitemShut {NoStop}%
\bibitem [{\citenamefont {Xue}\ \emph {et~al.}(2023)\citenamefont {Xue},
  \citenamefont {Sun}, \citenamefont {Wei}, \citenamefont {Li}, \citenamefont
  {Zhao}, \citenamefont {Wan}, \citenamefont {Lv}, \citenamefont {Zhao},
  \citenamefont {Xu},\ and\ \citenamefont {Li}}]{xuekun2023Generation}%
  \BibitemOpen
  \bibfield  {author} {\bibinfo {author} {\bibfnamefont {K.}~\bibnamefont
  {Xue}}, \bibinfo {author} {\bibfnamefont {T.}~\bibnamefont {Sun}}, \bibinfo
  {author} {\bibfnamefont {K.-J.}\ \bibnamefont {Wei}}, \bibinfo {author}
  {\bibfnamefont {Z.-P.}\ \bibnamefont {Li}}, \bibinfo {author} {\bibfnamefont
  {Q.}~\bibnamefont {Zhao}}, \bibinfo {author} {\bibfnamefont {F.}~\bibnamefont
  {Wan}}, \bibinfo {author} {\bibfnamefont {C.}~\bibnamefont {Lv}}, \bibinfo
  {author} {\bibfnamefont {Y.-T.}\ \bibnamefont {Zhao}}, \bibinfo {author}
  {\bibfnamefont {Z.-F.}\ \bibnamefont {Xu}},\ and\ \bibinfo {author}
  {\bibfnamefont {J.-X.}\ \bibnamefont {Li}},\ }\bibfield  {title} {\bibinfo
  {title} {{Generation of High-Density High-Polarization Positrons via
  Single-Shot Strong Laser-Foil Interaction}},\ }\href
  {https://doi.org/10.1103/PhysRevLett.131.175101} {\bibfield  {journal}
  {\bibinfo  {journal} {Phys. Rev. Lett.}\ }\textbf {\bibinfo {volume} {131}},\
  \bibinfo {pages} {175101} (\bibinfo {year} {2023})}\BibitemShut {NoStop}%
\bibitem [{\citenamefont {Ritus}(1985)}]{ritus1985quantum}%
  \BibitemOpen
  \bibfield  {author} {\bibinfo {author} {\bibfnamefont {V.}~\bibnamefont
  {Ritus}},\ }\bibfield  {title} {\bibinfo {title} {{Quantum effects of the
  interaction of elementary particles with an intense electromagnetic field}},\
  }\href {https://doi.org/10.1007/BF01120220} {\bibfield  {journal} {\bibinfo
  {journal} {J. Sov. Laser Res.}\ }\textbf {\bibinfo {volume} {6}},\ \bibinfo
  {pages} {497} (\bibinfo {year} {1985})}\BibitemShut {NoStop}%
\bibitem [{\citenamefont {Baier}\ \emph {et~al.}(1998)\citenamefont {Baier},
  \citenamefont {Katkov},\ and\ \citenamefont
  {Strakhovenko}}]{katkov1998electromagnetic}%
  \BibitemOpen
  \bibfield  {author} {\bibinfo {author} {\bibfnamefont {V.~N.}\ \bibnamefont
  {Baier}}, \bibinfo {author} {\bibfnamefont {V.~M.}\ \bibnamefont {Katkov}},\
  and\ \bibinfo {author} {\bibfnamefont {V.~M.}\ \bibnamefont {Strakhovenko}},\
  }\href {https://doi.org/10.1142/2216} {\emph {\bibinfo {title}
  {{Electromagnetic Processes at High Energies in Oriented Single Crystals}}}}\
  (\bibinfo  {publisher} {WORLD SCIENTIFIC},\ \bibinfo {year}
  {1998})\BibitemShut {NoStop}%
\bibitem [{\citenamefont {Di~Piazza}\ \emph {et~al.}(2019)\citenamefont
  {Di~Piazza}, \citenamefont {Tamburini}, \citenamefont {Meuren},\ and\
  \citenamefont {Keitel}}]{Piazza2019Improved}%
  \BibitemOpen
  \bibfield  {author} {\bibinfo {author} {\bibfnamefont {A.}~\bibnamefont
  {Di~Piazza}}, \bibinfo {author} {\bibfnamefont {M.}~\bibnamefont
  {Tamburini}}, \bibinfo {author} {\bibfnamefont {S.}~\bibnamefont {Meuren}},\
  and\ \bibinfo {author} {\bibfnamefont {C.~H.}\ \bibnamefont {Keitel}},\
  }\bibfield  {title} {\bibinfo {title} {{Improved local-constant-field
  approximation for strong-field QED codes}},\ }\href
  {https://doi.org/10.1103/PhysRevA.99.022125} {\bibfield  {journal} {\bibinfo
  {journal} {Phys. Rev. A}\ }\textbf {\bibinfo {volume} {99}},\ \bibinfo
  {pages} {022125} (\bibinfo {year} {2019})}\BibitemShut {NoStop}%
\bibitem [{\citenamefont {Lu}\ \emph {et~al.}(2007)\citenamefont {Lu},
  \citenamefont {Tzoufras}, \citenamefont {Joshi}, \citenamefont {Tsung},
  \citenamefont {Mori}, \citenamefont {Vieira}, \citenamefont {Fonseca},\ and\
  \citenamefont {Silva}}]{lu2007generating}%
  \BibitemOpen
  \bibfield  {author} {\bibinfo {author} {\bibfnamefont {W.}~\bibnamefont
  {Lu}}, \bibinfo {author} {\bibfnamefont {M.}~\bibnamefont {Tzoufras}},
  \bibinfo {author} {\bibfnamefont {C.}~\bibnamefont {Joshi}}, \bibinfo
  {author} {\bibfnamefont {F.~S.}\ \bibnamefont {Tsung}}, \bibinfo {author}
  {\bibfnamefont {W.~B.}\ \bibnamefont {Mori}}, \bibinfo {author}
  {\bibfnamefont {J.}~\bibnamefont {Vieira}}, \bibinfo {author} {\bibfnamefont
  {R.~A.}\ \bibnamefont {Fonseca}},\ and\ \bibinfo {author} {\bibfnamefont
  {L.~O.}\ \bibnamefont {Silva}},\ }\bibfield  {title} {\bibinfo {title}
  {{Generating multi-{G}eV electron bunches using single stage laser wakefield
  acceleration in a 3{D} nonlinear regime}},\ }\href
  {https://doi.org/10.1103/PhysRevSTAB.10.061301} {\bibfield  {journal}
  {\bibinfo  {journal} {Phys. Rev. ST Accel. Beams}\ }\textbf {\bibinfo
  {volume} {10}},\ \bibinfo {pages} {061301} (\bibinfo {year}
  {2007})}\BibitemShut {NoStop}%
\bibitem [{\citenamefont {Leemans}\ \emph {et~al.}(2014)\citenamefont
  {Leemans}, \citenamefont {Gonsalves}, \citenamefont {Mao}, \citenamefont
  {Nakamura}, \citenamefont {Benedetti}, \citenamefont {Schroeder},
  \citenamefont {T\'oth}, \citenamefont {Daniels}, \citenamefont
  {Mittelberger}, \citenamefont {Bulanov}, \citenamefont {Vay}, \citenamefont
  {Geddes},\ and\ \citenamefont {Esarey}}]{leemans2014multi}%
  \BibitemOpen
  \bibfield  {author} {\bibinfo {author} {\bibfnamefont {W.~P.}\ \bibnamefont
  {Leemans}}, \bibinfo {author} {\bibfnamefont {A.~J.}\ \bibnamefont
  {Gonsalves}}, \bibinfo {author} {\bibfnamefont {H.-S.}\ \bibnamefont {Mao}},
  \bibinfo {author} {\bibfnamefont {K.}~\bibnamefont {Nakamura}}, \bibinfo
  {author} {\bibfnamefont {C.}~\bibnamefont {Benedetti}}, \bibinfo {author}
  {\bibfnamefont {C.~B.}\ \bibnamefont {Schroeder}}, \bibinfo {author}
  {\bibfnamefont {C.}~\bibnamefont {T\'oth}}, \bibinfo {author} {\bibfnamefont
  {J.}~\bibnamefont {Daniels}}, \bibinfo {author} {\bibfnamefont {D.~E.}\
  \bibnamefont {Mittelberger}}, \bibinfo {author} {\bibfnamefont {S.~S.}\
  \bibnamefont {Bulanov}}, \bibinfo {author} {\bibfnamefont {J.-L.}\
  \bibnamefont {Vay}}, \bibinfo {author} {\bibfnamefont {C.~G.~R.}\
  \bibnamefont {Geddes}},\ and\ \bibinfo {author} {\bibfnamefont
  {E.}~\bibnamefont {Esarey}},\ }\bibfield  {title} {\bibinfo {title}
  {{Multi-{G}eV Electron Beams from Capillary-Discharge-Guided Subpetawatt
  Laser Pulses in the Self-Trapping Regime}},\ }\href
  {https://doi.org/10.1103/PhysRevLett.113.245002} {\bibfield  {journal}
  {\bibinfo  {journal} {Phys. Rev. Lett.}\ }\textbf {\bibinfo {volume} {113}},\
  \bibinfo {pages} {245002} (\bibinfo {year} {2014})}\BibitemShut {NoStop}%
\bibitem [{\citenamefont {Babjak}\ \emph {et~al.}(2024)\citenamefont {Babjak},
  \citenamefont {Willingale}, \citenamefont {Arefiev},\ and\ \citenamefont
  {Vranic}}]{Babjak2024Direct}%
  \BibitemOpen
  \bibfield  {author} {\bibinfo {author} {\bibfnamefont {R.}~\bibnamefont
  {Babjak}}, \bibinfo {author} {\bibfnamefont {L.}~\bibnamefont {Willingale}},
  \bibinfo {author} {\bibfnamefont {A.}~\bibnamefont {Arefiev}},\ and\ \bibinfo
  {author} {\bibfnamefont {M.}~\bibnamefont {Vranic}},\ }\bibfield  {title}
  {\bibinfo {title} {{Direct Laser Acceleration in Underdense Plasmas with
  Multi-PW Lasers: A Path to High-Charge, GeV-Class Electron Bunches}},\ }\href
  {https://doi.org/10.1103/PhysRevLett.132.125001} {\bibfield  {journal}
  {\bibinfo  {journal} {Phys. Rev. Lett.}\ }\textbf {\bibinfo {volume} {132}},\
  \bibinfo {pages} {125001} (\bibinfo {year} {2024})}\BibitemShut {NoStop}%
\bibitem [{\citenamefont {LEE}\ \emph {et~al.}(1998)\citenamefont {LEE},
  \citenamefont {FUKAHORI},\ and\ \citenamefont {CHANG}}]{Young1998Evaluation}%
  \BibitemOpen
  \bibfield  {author} {\bibinfo {author} {\bibfnamefont {Y.-O.}\ \bibnamefont
  {LEE}}, \bibinfo {author} {\bibfnamefont {T.}~\bibnamefont {FUKAHORI}},\ and\
  \bibinfo {author} {\bibfnamefont {J.}~\bibnamefont {CHANG}},\ }\bibfield
  {title} {\bibinfo {title} {{Evaluation of Photonuclear Reaction Data on
  Tantalum-181 up to 140MeV}},\ }\href
  {https://doi.org/10.1080/18811248.1998.9733928} {\bibfield  {journal}
  {\bibinfo  {journal} {J. Nucl. Sci. Technol.}\ }\textbf {\bibinfo {volume}
  {35}},\ \bibinfo {pages} {685} (\bibinfo {year} {1998})}\BibitemShut
  {NoStop}%
\bibitem [{qua(2000)}]{quasi-deuteron-international2000iaea}%
  \BibitemOpen
  \href
  {https://www.iaea.org/publications/6043/handbook-on-photonuclear-data-for-applications-cross-sections-and-spectra}
  {\emph {\bibinfo {title} {{Handbook on Photonuclear Data for Applications
  Cross-sections and Spectra}}}},\ \bibinfo {series} {TECDOC Series}\ No.\
  \bibinfo {number} {1178}\ (\bibinfo  {publisher} {INTERNATIONAL ATOMIC ENERGY
  AGENCY},\ \bibinfo {address} {Vienna},\ \bibinfo {year} {2000})\BibitemShut
  {NoStop}%
\bibitem [{\citenamefont {Chakrabarty}\ \emph {et~al.}(2016)\citenamefont
  {Chakrabarty}, \citenamefont {Dinh~Dang},\ and\ \citenamefont
  {Datar}}]{Chakrabarty2016GDR}%
  \BibitemOpen
  \bibfield  {author} {\bibinfo {author} {\bibfnamefont {D.~R.}\ \bibnamefont
  {Chakrabarty}}, \bibinfo {author} {\bibfnamefont {N.}~\bibnamefont
  {Dinh~Dang}},\ and\ \bibinfo {author} {\bibfnamefont {V.~M.}\ \bibnamefont
  {Datar}},\ }\bibfield  {title} {\bibinfo {title} {Giant dipole resonance in
  hot rotating nuclei},\ }\href {https://doi.org/10.1140/epja/i2016-16143-9}
  {\bibfield  {journal} {\bibinfo  {journal} {Eur. Phys. J. A}\ }\textbf
  {\bibinfo {volume} {52}},\ \bibinfo {pages} {143} (\bibinfo {year}
  {2016})}\BibitemShut {NoStop}%
\bibitem [{\citenamefont {Katsuma}(2014)}]{Katsuma2014Photoelectric}%
  \BibitemOpen
  \bibfield  {author} {\bibinfo {author} {\bibfnamefont {M.}~\bibnamefont
  {Katsuma}},\ }\bibfield  {title} {\bibinfo {title} {Photoelectric
  disintegration of $^{16}\mathrm{O}$},\ }\href
  {https://doi.org/10.1103/PhysRevC.90.068801} {\bibfield  {journal} {\bibinfo
  {journal} {Phys. Rev. C}\ }\textbf {\bibinfo {volume} {90}},\ \bibinfo
  {pages} {068801} (\bibinfo {year} {2014})}\BibitemShut {NoStop}%
\bibitem [{\citenamefont {Downer}\ \emph {et~al.}(2018)\citenamefont {Downer},
  \citenamefont {Zgadzaj}, \citenamefont {Debus}, \citenamefont {Schramm},\
  and\ \citenamefont {Kaluza}}]{Downer2018iagnostics}%
  \BibitemOpen
  \bibfield  {author} {\bibinfo {author} {\bibfnamefont {M.~C.}\ \bibnamefont
  {Downer}}, \bibinfo {author} {\bibfnamefont {R.}~\bibnamefont {Zgadzaj}},
  \bibinfo {author} {\bibfnamefont {A.}~\bibnamefont {Debus}}, \bibinfo
  {author} {\bibfnamefont {U.}~\bibnamefont {Schramm}},\ and\ \bibinfo {author}
  {\bibfnamefont {M.~C.}\ \bibnamefont {Kaluza}},\ }\bibfield  {title}
  {\bibinfo {title} {Diagnostics for plasma-based electron accelerators},\
  }\href {https://doi.org/10.1103/RevModPhys.90.035002} {\bibfield  {journal}
  {\bibinfo  {journal} {Rev. Mod. Phys.}\ }\textbf {\bibinfo {volume} {90}},\
  \bibinfo {pages} {035002} (\bibinfo {year} {2018})}\BibitemShut {NoStop}%
\bibitem [{\citenamefont {Tajima}\ and\ \citenamefont
  {Dawson}(1979)}]{Tajima1979Laser}%
  \BibitemOpen
  \bibfield  {author} {\bibinfo {author} {\bibfnamefont {T.}~\bibnamefont
  {Tajima}}\ and\ \bibinfo {author} {\bibfnamefont {J.~M.}\ \bibnamefont
  {Dawson}},\ }\bibfield  {title} {\bibinfo {title} {{Laser Electron
  Accelerator}},\ }\href {https://doi.org/10.1103/PhysRevLett.43.267}
  {\bibfield  {journal} {\bibinfo  {journal} {Phys. Rev. Lett.}\ }\textbf
  {\bibinfo {volume} {43}},\ \bibinfo {pages} {267} (\bibinfo {year}
  {1979})}\BibitemShut {NoStop}%
\bibitem [{\citenamefont {Esarey}\ \emph {et~al.}(2009)\citenamefont {Esarey},
  \citenamefont {Schroeder},\ and\ \citenamefont
  {Leemans}}]{Leemans2009Physics}%
  \BibitemOpen
  \bibfield  {author} {\bibinfo {author} {\bibfnamefont {E.}~\bibnamefont
  {Esarey}}, \bibinfo {author} {\bibfnamefont {C.~B.}\ \bibnamefont
  {Schroeder}},\ and\ \bibinfo {author} {\bibfnamefont {W.~P.}\ \bibnamefont
  {Leemans}},\ }\bibfield  {title} {\bibinfo {title} {Physics of laser-driven
  plasma-based electron accelerators},\ }\href
  {https://doi.org/10.1103/RevModPhys.81.1229} {\bibfield  {journal} {\bibinfo
  {journal} {Rev. Mod. Phys.}\ }\textbf {\bibinfo {volume} {81}},\ \bibinfo
  {pages} {1229} (\bibinfo {year} {2009})}\BibitemShut {NoStop}%
\bibitem [{\citenamefont {Schroeder}\ \emph {et~al.}(2011)\citenamefont
  {Schroeder}, \citenamefont {Benedetti}, \citenamefont {Esarey}, \citenamefont
  {Gr\"uner},\ and\ \citenamefont {Leemans}}]{Schroeder2011Growth}%
  \BibitemOpen
  \bibfield  {author} {\bibinfo {author} {\bibfnamefont {C.~B.}\ \bibnamefont
  {Schroeder}}, \bibinfo {author} {\bibfnamefont {C.}~\bibnamefont
  {Benedetti}}, \bibinfo {author} {\bibfnamefont {E.}~\bibnamefont {Esarey}},
  \bibinfo {author} {\bibfnamefont {F.~J.}\ \bibnamefont {Gr\"uner}},\ and\
  \bibinfo {author} {\bibfnamefont {W.~P.}\ \bibnamefont {Leemans}},\
  }\bibfield  {title} {\bibinfo {title} {{Growth and Phase Velocity of
  Self-Modulated Beam-Driven Plasma Waves}},\ }\href
  {https://doi.org/10.1103/PhysRevLett.107.145002} {\bibfield  {journal}
  {\bibinfo  {journal} {Phys. Rev. Lett.}\ }\textbf {\bibinfo {volume} {107}},\
  \bibinfo {pages} {145002} (\bibinfo {year} {2011})}\BibitemShut {NoStop}%
\bibitem [{\citenamefont {Pukhov}\ \emph {et~al.}(2011)\citenamefont {Pukhov},
  \citenamefont {Kumar}, \citenamefont {T\"uckmantel}, \citenamefont
  {Upadhyay}, \citenamefont {Lotov}, \citenamefont {Muggli}, \citenamefont
  {Khudik}, \citenamefont {Siemon},\ and\ \citenamefont
  {Shvets}}]{Pukhov2011Phase}%
  \BibitemOpen
  \bibfield  {author} {\bibinfo {author} {\bibfnamefont {A.}~\bibnamefont
  {Pukhov}}, \bibinfo {author} {\bibfnamefont {N.}~\bibnamefont {Kumar}},
  \bibinfo {author} {\bibfnamefont {T.}~\bibnamefont {T\"uckmantel}}, \bibinfo
  {author} {\bibfnamefont {A.}~\bibnamefont {Upadhyay}}, \bibinfo {author}
  {\bibfnamefont {K.}~\bibnamefont {Lotov}}, \bibinfo {author} {\bibfnamefont
  {P.}~\bibnamefont {Muggli}}, \bibinfo {author} {\bibfnamefont
  {V.}~\bibnamefont {Khudik}}, \bibinfo {author} {\bibfnamefont
  {C.}~\bibnamefont {Siemon}},\ and\ \bibinfo {author} {\bibfnamefont
  {G.}~\bibnamefont {Shvets}},\ }\bibfield  {title} {\bibinfo {title} {{Phase
  Velocity and Particle Injection in a Self-Modulated Proton-Driven Plasma
  Wakefield Accelerator}},\ }\href
  {https://doi.org/10.1103/PhysRevLett.107.145003} {\bibfield  {journal}
  {\bibinfo  {journal} {Phys. Rev. Lett.}\ }\textbf {\bibinfo {volume} {107}},\
  \bibinfo {pages} {145003} (\bibinfo {year} {2011})}\BibitemShut {NoStop}%
\bibitem [{\citenamefont {Whittum}\ \emph {et~al.}(1991)\citenamefont
  {Whittum}, \citenamefont {Sharp}, \citenamefont {Yu}, \citenamefont {Lampe},\
  and\ \citenamefont {Joyce}}]{Whittum1991Electron}%
  \BibitemOpen
  \bibfield  {author} {\bibinfo {author} {\bibfnamefont {D.~H.}\ \bibnamefont
  {Whittum}}, \bibinfo {author} {\bibfnamefont {W.~M.}\ \bibnamefont {Sharp}},
  \bibinfo {author} {\bibfnamefont {S.~S.}\ \bibnamefont {Yu}}, \bibinfo
  {author} {\bibfnamefont {M.}~\bibnamefont {Lampe}},\ and\ \bibinfo {author}
  {\bibfnamefont {G.}~\bibnamefont {Joyce}},\ }\bibfield  {title} {\bibinfo
  {title} {Electron-hose instability in the ion-focused regime},\ }\href
  {https://doi.org/10.1103/PhysRevLett.67.991} {\bibfield  {journal} {\bibinfo
  {journal} {Phys. Rev. Lett.}\ }\textbf {\bibinfo {volume} {67}},\ \bibinfo
  {pages} {991} (\bibinfo {year} {1991})}\BibitemShut {NoStop}%
\bibitem [{\citenamefont {Schroeder}\ \emph {et~al.}(2012)\citenamefont
  {Schroeder}, \citenamefont {Benedetti}, \citenamefont {Esarey}, \citenamefont
  {Gr\"uner},\ and\ \citenamefont {Leemans}}]{Schroeder2012Coupled}%
  \BibitemOpen
  \bibfield  {author} {\bibinfo {author} {\bibfnamefont {C.~B.}\ \bibnamefont
  {Schroeder}}, \bibinfo {author} {\bibfnamefont {C.}~\bibnamefont
  {Benedetti}}, \bibinfo {author} {\bibfnamefont {E.}~\bibnamefont {Esarey}},
  \bibinfo {author} {\bibfnamefont {F.~J.}\ \bibnamefont {Gr\"uner}},\ and\
  \bibinfo {author} {\bibfnamefont {W.~P.}\ \bibnamefont {Leemans}},\
  }\bibfield  {title} {\bibinfo {title} {Coupled beam hose and self-modulation
  instabilities in overdense plasma},\ }\href
  {https://doi.org/10.1103/PhysRevE.86.026402} {\bibfield  {journal} {\bibinfo
  {journal} {Phys. Rev. E}\ }\textbf {\bibinfo {volume} {86}},\ \bibinfo
  {pages} {026402} (\bibinfo {year} {2012})}\BibitemShut {NoStop}%
\end{thebibliography}%
\end{document}